\documentclass[journal]{IEEEtran}
\usepackage{amsmath,amsfonts}
\usepackage{algorithmic}
\usepackage{array}
\usepackage[caption=false,font=normalsize,labelfont=sf,textfont=sf]{subfig}
\usepackage{textcomp}
\usepackage{stfloats}
\usepackage{booktabs}
\usepackage{url}
\usepackage{verbatim}
\usepackage{graphicx}
\usepackage{booktabs}
\usepackage{threeparttable}
\def\BibTeX{{\rm B\kern-.05em{\sc i\kern-.025em b}\kern-.08em
    T\kern-.1667em\lower.7ex\hbox{E}\kern-.125emX}}
\usepackage{balance}
\usepackage{multirow}
\usepackage{footnote}

\usepackage[table,xcdraw]{xcolor}
\usepackage[numbers,sort&compress]{natbib}

\DeclareMathOperator{\tr}{tr}
\newtheorem{definition}{Definition}[section]
\usepackage{arydshln}
\begin{document}
\title{BrainIB: Interpretable Brain Network-based Psychiatric Diagnosis with Graph Information Bottleneck}
\author{Kaizhong Zheng, Shujian Yu, Baojuan Li, Robert Jenssen, and Badong Chen
\thanks{This work was supported in part by the National Natural Science Foundation of China with grant numbers (62088102, U21A20485, 62311540022), and the Research Council of Norway with grant number 309439. \emph{(Corresponding authors: Shujian Yu; Badong Chen.)}

Kaizhong Zheng and Badong Chen are with National Key Laboratory of Human-Machine Hybrid Augmented Intelligence, National Engineering Research Center for Visual Information and Applications, and Institute of Artificial Intelligence and Robotics, Xi’an Jiaotong University, Xi'an (Email: kzzheng@stu.xjtu.edu.cn; chenbd@mail.xjtu.edu.cn).

Shujian Yu is with the Department of Computer Science, Vrije Universiteit Amsterdam, Amsterdam and the Machine Learning Group, UiT - Arctic University of Norway, Troms{\o} (Email: yusj9011@gmail.com)

Robert Jenssen is with the Machine Learning Group, UiT - Arctic University of Norway, Troms{\o} (Email: robert.jenssen@uit.no).

Baojuan Li is with the School of Biomedical Engineering, Fourth Military Medical University, Xi’an (Email: libjuan@163.com).

}}

\markboth{}%
{}

\maketitle

\begin{abstract}
Developing a new diagnostic models based on the underlying biological mechanisms rather than subjective symptoms for psychiatric disorders is an emerging consensus. Recently, machine learning-based classifiers using functional connectivity (FC) for psychiatric disorders and healthy controls are developed to identify brain markers. However, existing machine learning-based diagnostic models are prone to over-fitting (due to insufficient training samples) and  perform poorly in new test environment. Furthermore, it is difficult to obtain explainable and reliable brain biomarkers elucidating the underlying diagnostic decisions. These issues hinder their possible clinical applications. In this work, we propose BrainIB, a new graph neural network (GNN) framework to analyze functional magnetic resonance images (fMRI), by leveraging the famed Information Bottleneck (IB) principle. BrainIB is able to identify the most informative edges in the brain (i.e., subgraph) and generalizes well to unseen data. We evaluate the performance of BrainIB against $3$ baselines and $7$ state-of-the-art brain network classification methods on three psychiatric datasets and observe that our BrainIB always achieves the highest diagnosis accuracy. It also discovers the subgraph biomarkers which are consistent to clinical and neuroimaging findings. The source code and implementation details of BrainIB are freely available at GitHub repository (https://github.com/SJYuCNEL/brain-and-Information-Bottleneck/).

\end{abstract}

\begin{IEEEkeywords}
Psychiatric diagnosis, graph neural network (GNN), Information bottleneck, brain network, functional magnetic resonance imaging (fMRI).
\end{IEEEkeywords}

\section{Introduction}
\IEEEPARstart{P}{sychiatric} disorders (such as depression and autism) are the leading causes of disability worldwide~\cite{marshall2020hidden}, whereas psychiatric diagnoses remain a challenging open issue. Existing clinical diagnosis of psychiatric disorders relies heavily on constellations of symptoms \cite{zhang2021identification}, such as emotional, cognitive symptoms \emph{etc}. In general, patients with psychiatric disorders are diagnosed by psychiatrists using criteria from the Diagnostic and Statistical Manual of Mental Disorders $5$th edition (DSM-V) \cite{american1994american}. However, traditional symptom-based diagnosis is insufficient and may lead to misdiagnosis, due to the clinical heterogeneity \cite{goodkind2015identification}. Therefore, developing an effective diagnostic tool based on the underlying biological mechanisms rather than symptoms is an emerging consensus.

Functional magnetic resonance imaging (fMRI) \cite{matthews2004functional} is a noninvasive neuroimaging technique that has been widely used to characterize the underlying pathophysiology of psychiatric disorders. The resting-state fMRI (rs-fMRI) can be used to study and assess alternations of whole brain functional connectivity (FC) network in diverse patient populations~\cite{biswal2010toward,xia2017functional}. Previous studies demonstrated that fMRI-based characterizations are reliable complements to the existing symptom-based diagnoses~\cite{biswal2010toward,xia2017functional}.

Machine learning (ML) techniques that use fMRI FC measures as input have been extensively investigated for psychiatric diagnosis. Earlier methods use shallow or simple classification models such as support vector machines (SVM)~\cite{pan2018novel} and random forest (RF)~\cite{ronicko2020diagnostic}, which are incapable of analyzing nonlinear information of brain network. Deep neural networks have gained popularity in recent years due to their strong representation power \cite{li2021braingnn}.  






In general, brain networks can be viewed as complex graphs with anatomic brain regions of interest (ROIs) represented as nodes and FC between brain ROIs as edges. This motivates the applications of graph neural networks (GNNs)~\cite{kipf2016semi} for psychiatric diagnosis~\cite{rakhimberdina2020population,rakhimberdina2019linear}. So far, GNNs have achieved promising diagnostic accuracy on autism spectrum disorder (ASD)~\cite{rakhimberdina2020population} and schizophrenia~\cite{rakhimberdina2019linear}.


Despite of recent performance gains, existing ML-based diagnostic models still suffer from the following issues: 

\begin{enumerate}
\item{\textbf{Interpretability:} Most of existing diagnostic models~\cite{rakhimberdina2020population, yao2019triplet} cannot discover explainable brain biomarkers (e.g., ROIs as groups of nodes or edges) elucidating the underlying diagnostic decisions and revealing neural mechanism of the disease.}

\item{\textbf{Generalization:} Most of existing diagnostic models are trained in a homogeneous or single site dataset with a small number of samples. This may increase the probability of over-fitting and lead to poor generalization capacity during deployment. For example, \cite{zeng2012identifying} only considers $24$ patients with major depression. 
}
\end{enumerate}

 The information bottleneck (IB) principle~\cite{tishby1999information} in information theory holds promise for addressing the aforementioned issues. IB principle aims to extract a \emph{compact} representations from the original data which is most \emph{informative} to label. By eliminating redundant information, this approach can significantly enhance model generalization and improve interpretability~\cite{shamir2010learning}. Recently, the general idea of IB has recently been extended to GNNs. However, current graph IB methods, like subgraph information bottleneck (SIB)~\cite{yu2021recognizing}, identify important subgraphs based on node selection rather than edge selection. In the diagnosis of psychiatric disorders, brain connectivity (edge) is more important than brain regions (node) for two reasons. Firstly, mounting evidence indicates that brain connectivity plays a crucial role in the diagnosis and understanding of the neural mechanisms underlying psychiatric disorders~\cite{insel2015brain,lozano2017waving}. For instance, Insel et al.~\cite{insel2015brain} in the journal Science posited that future diagnostics may redefine ``mental disorders" as ``brain circuit disorders". Secondly, numerous studies on psychiatric diagnosis have identified potential biomarkers linked to brain connectivity~\cite{gallo2023functional,li2020neuroimaging}. Furthermore, current graph IB approaches have only been validated on small-scale datasets (comprising dozens of nodes and edges), such as those involving molecule~\cite{debnath1991structure}. Their performance and applicability on large-scale datasets, including those related to brain diseases (involving hundreds of nodes and thousands of edges), remain unknown.

In this paper,  we introduce the information bottleneck principle to brain network analysis and develop a new GNN-based intepretable brain network classification framework that is able to identify the most informative subgraph to the decision and generalizes well to unseen data. We term our framework the brain information bottleneck (BrainIB) and evaluate it in two multi-site, large-scale datasets.


 To summarize, our contributions entail novel advancements in methodology, generalization, interpretability:
\begin{itemize}
\item  In terms of methodology, our BrainIB makes two improvements over current graph IB approaches:
\begin{itemize}
    \item Instead of using the mutual information neural estimator (MINE)~\cite{belghazi2018mutual}, the matrix-based R{\'e}nyi's $\alpha$-order entropy functional~\cite{yu2019multivariate} is used to measure mutual information values in graph information bottleneck, which significantly stabilizes the training ( see Section~\ref{stable} Stable Training Discussion).
    \item We optimize subgraph generator specifically for brain network analysis, which is able to discover informative edges. 
    
\end{itemize}

\item{In terms of generalization capability, we use BrainIB against 3 baselines and 7 popular brain network classification methods (including SIB) on two multi-site, large-scale datasets, i.e., ABIDE \cite{di2014autism} and REST-meta-MDD \cite{yan2019reduced}.  ABIDE contains 1099 participants (528 ASDs and 571 typically developed individuals) from 17 sites and REST-meta-MDD includes 1604 participants (848 MDDs and 794 healthy controls) from 17 hospitals in China. Our BrainIB demonstrates overwhelming performance gain in both $10$-fold and leave-one-site-out cross validations.}


\item{ In terms of interpretability, we obtain disease-specific prominent brain network connections/systems in patients with major depressive disorder (MDD), autism spectrum disorder (ASD) and schizophrenia. Our discovered biomarkers are consistent with clinical and neuroimaging findings.}

\end{itemize}

 The remaining of this paper is organized as follows. Section II shows contextual terms and approaches used in the paper. Section III briefly introduces the related work. Section IV elaborates our BrainIB which consists of three modules: subgraph generator, graph encoder and mutual information estimation module. Experiments results are presented in Section V and VI. Finally, Section VII draws the conclusion.

\section{Preliminaries}
Table~\ref{tab:notations} demonstrates contextual terms and approaches used in the paper.
\begin{table}[ht!]
\centering
\caption{  Contextual terms and approaches used in the paper}\label{tab:notations}
\resizebox{\linewidth}{!}{\begin{tabular}{@{}ll@{}}
\toprule
\textbf{Notations} & \textbf{Description} \\ \midrule
\text{$\text{ABIDE~\cite{di2014autism}}$}          & Autism Brain Imaging Data Exchange I\\
\text{$\text{REST-meta-MDD~\cite{yan2019reduced}}$}          & The largest resting-state fMRI dataset for major depression\\
\text{$\text{SRPBS~\cite{tanaka2021multi}}$}          &Multi-disorder MRI dataset\\ 
\text{$\text{SVM~\cite{2015Functional}}$}          &  Support vector machines\\ 
\text{$\text{GCN~\cite{kipf2016semi}}$}          & Graph convolutional network\\ 
\text{$\text{GAT~\cite{velivckovic2018graph}}$}          & Graph attention network\\ 
\text{$\text{GIN~\cite{xu2018powerful}}$}          & Graph isomorphism network\\
\text{$\text{SIB~\cite{yu2021recognizing}}$}          & Subgraph information bottleneck\\
\text{$\text{VGIB~\cite{yu2022improving}}$}          & Variational graph information bottleneck \\
\text{$\text{DIR-GNN~\cite{wudiscovering}}$}          & Discovering invariant rationales for graph neural networks \\
\text{$\text{ProtGNN~\cite{zhang2022protgnn}}$}          & Prototype graph neural network\\
\text{$\text{BrainGNN~\cite{li2021braingnn}}$}          & Interpretable brain graph neural network\\
\text{$\text{IBGNN~\cite{cui2022interpretable}}$}          & Interpretable graph neural networks for brain disorder analysis\\
\text{$\text{CI-GNN~\cite{zheng2024ci}}$}          & Causality-inspired graph neural network
\\\bottomrule
\end{tabular}}
\end{table}

\section{Related work}
\subsection{Diagnostic Models For Psychiatric Disorders}

The identification of predictive subnetworks and edges is an essential procedure for the development of modern psychiatric diagnostic models~\cite{wang2021learning}. 
Traditionally, this is done by treating functional connectivities as features and performing feature selection to preserve the most salient connections. Popular feature selection methods include statistical test~\cite{du2018classification} like t-test or ranksum-test and LASSO. Our BrainIB is an end-to-end disease diagnostic model that is able to remove irrelevant or less-informative edges without an explicit feature selection procedure.


Earlier classification models for psychiatric disorders include support vector machines (SVM) and random forest (RF)~\cite{pan2018novel}. For example, \cite{2015Functional} uses linear SVM to discriminate autism patients from healthy controls and achieves an overall accuracy of $0.697$.  However, these shallow learning methods could not capture topological information within  complex brain network structures \cite{huang2020identifying} and achieve the acceptable performance on the large-scale data sets. The most recent studies resort to GNNs to further improve diagnostic accuracy. For example, \cite{parisot2018disease} applies graph convolutional networks (GCNs) also on autism dataset and obtain a higher accuracy of 0.704. Despite the obvious performance gain, GNNs are always ``black-box" algorithms , which makes it hard to understand their decision making process - a major issue for clinical applications. In this study, we propose a \emph{built-in} explainable diagnostic model which enables automatically recognize subgraphs elucidating the underlying diagnostic decisions. 




\subsection{Graph Neural Networks For Graph Classification}

Brain networks are complex graphs in which anatomic regions denote nodes and functional connectivities denote edges. Therefore, psychiatric diagnosis can be regarded as graph classification task. Given  an input graph $\mathcal{G}=(\mathcal{V},\mathcal{E})$ with node feature matrix $X$, GNNs employ the message-passing paradigm to propagate and aggregate the representations of information along edges to generate a node representation $h_{v}$ for each node $v \in \mathcal{V}$. Formally, a GNN can be defined through an aggregation function $\mbox{A}$ and a combine function $\mbox{C}$ such that for the $k$-th layer:
\begin{equation}
a_{v}^{\left ( k \right )}=\mbox{A}^{\left ( k \right )}\left ( \left\{ h_{u}^{\left ( k-1 \right )} :u \in \mathcal{N} \left ( v \right )\right\} \right ),
\end{equation}

\begin{equation}
h_{v}^{\left ( k \right )}= \mbox{C}^{\left ( k \right )}\left ( h_{v}^{\left ( k-1 \right )},a_{v}^{\left ( k \right )} \right ),
\end{equation}
where $h_{v}^{\left ( k \right )}$ is the node embedding of node $v$ at the $k$-th layer and $\mathcal{N} \left ( v \right )$ is the set of neighbour nodes of $v$. In general, the aggregation strategies of GNN include mean- \cite{kipf2016semi}, sum- \cite{xu2018powerful}, or max-pooling \cite{hamilton2017inductive} . Here, we use Graph Isomorphism Networks (GIN) which uses sum-pooling as the aggregation strategy, as an example. Its message passing procedures:

\begin{equation}
h_{v}^{k}=\mbox{MLP}^{k}\left ( \left ( 1+ \epsilon^{k} \right )\cdot h_{v}^{k-1} + \sum_{u\in\mathcal{N}\left (  v\right )}^{}h_{u}^{k-1}\right )
\end{equation}
where $\mbox{MLP}$ is the multi-layer perceptron, $\epsilon$ refers to a learnable parameter. We initialize  $H^{0}=X$ in the first iteration.


For graph classification task, the entire graph’s representation $h_{G}$ is obtained from node embedding $h_{v}^{k}$ through the READOUT function $\mbox{R}$:
\begin{equation}
h_{G}=\mbox{R}\left ( \left\{h_{v}^{\left ( k \right )}|v \in G \right\} \right ).
\end{equation}

Averaging and summation~\cite{kipf2016semi,xu2018powerful} are the most common strategies for the READOUT function. Another popular strategy is the hierarchical graph pooling~\cite{ying2018hierarchical} that decreases the number of nodes to one. 



\subsection{Information Bottleneck and GNN Interpretability}
In a typical learning scenario and more specifically classification tasks, we have input $X$ and corresponding desired output $Y$, and we seek to find a mapping between $X$ and $Y$ via observing a finite sample generated by a fixed but unknown distribution $p(x,y)$. The Information Bottleneck (IB) principle~\cite{tishby1999information} formulates the learning as:
\begin{equation}
    \min_{p(t|x)} I(X;T)-\beta I(Y;T),
\end{equation}
in which $I(\cdot;\cdot)$ denotes mutual information, $T$ is the latent representation of the input $X$. $\beta>0$ is a Lagrange multiplier that controls the trade-off between the \textbf{minimality} or complexity of the representation (as measured by $I(X;T)$) and the \textbf{sufficiency} of the representation $T$ to the performance of the task (as quantified by $I(Y;T)$). In this sense, the IB principle also provides a natural approximation of \emph{minimal sufficient statistic}~\cite{gilad2003information}.

The general idea of IB has recently been extended to GNNs. Let $\mathcal{G}$ denote graph input data which encodes both graph structure information (characterized by either adjacency matrix $A$) and node attribute matrix $X$, and $Y$ the desired response such as node labels or graph labels. The Subgraph Information Bottleneck (SIB)~\cite{yu2021recognizing} aims to extract the most informative or interpretable subgraph $\mathcal{G}_{\text{sub}}$ from $\mathcal{G}$ by the following objective:
\begin{equation}\label{eq:GIB_Lagrangian}
    \mathcal{L}_{\text{SIB}}=\min I(\mathcal{G};\mathcal{G}_{\text{sub}}) - \beta I(Y;\mathcal{G}_{\text{sub}}).
\end{equation}

 Yu \emph{et al} \cite{yu2021recognizing} approximates $-I(Y;G_{\text{sub}})$ by minimizing the cross-entropy loss and extracts the subgraph by removing redundant or irrelevant nodes. The mutual information term $I(\mathcal{G};\mathcal{G}_{\text{sub}})$ is evaluated by the mutual information neural estimator (MINE)~\cite{belghazi2018mutual} which requires an additional network and is highly unstable during training. In another parallel work, the IB principle has been used to learn compressed node representations~\cite{wu2020graph}.

The SIB can be viewed as a \emph{built-in} interpretable GNNs (i.e. self-explaining GNNs), as it can automatically identify the informative subgraph that is mostly influential to decision or graph label $Y$. The GNN interpretability has recently gained increased attention. We refer interested readers to a recent survey~\cite{yuan2020explainability} on this topic. However, most of existing interpretation methods are \emph{post-hoc}, which means another explanatory model is used to provide explanations for a well-trained GNN. Notable examples include~\cite{luo2020parameterized,yuan2021explainability}.  It remains a question that the \emph{post-hoc} explanation is unreliable in the process of underlying diagnostic decision compared with self-explaining methods \cite{rudin2019stop}. In the application of brain network classification, BrainNNExplainer~\cite{cui2021brainnnexplainer}, learns a global mask to highlight disease-specific prominent brain network connections, whereas BrainGNN~\cite{li2021braingnn} designed region-of-interest (ROI) aware graph convolutional layer and pooling layer to highlight salient ROIs (nodes in the graph).

\section{Proposed framework}

\subsection{Problem Definition}

\begin{figure*}[ht!]
\centering
\includegraphics[scale=0.58]{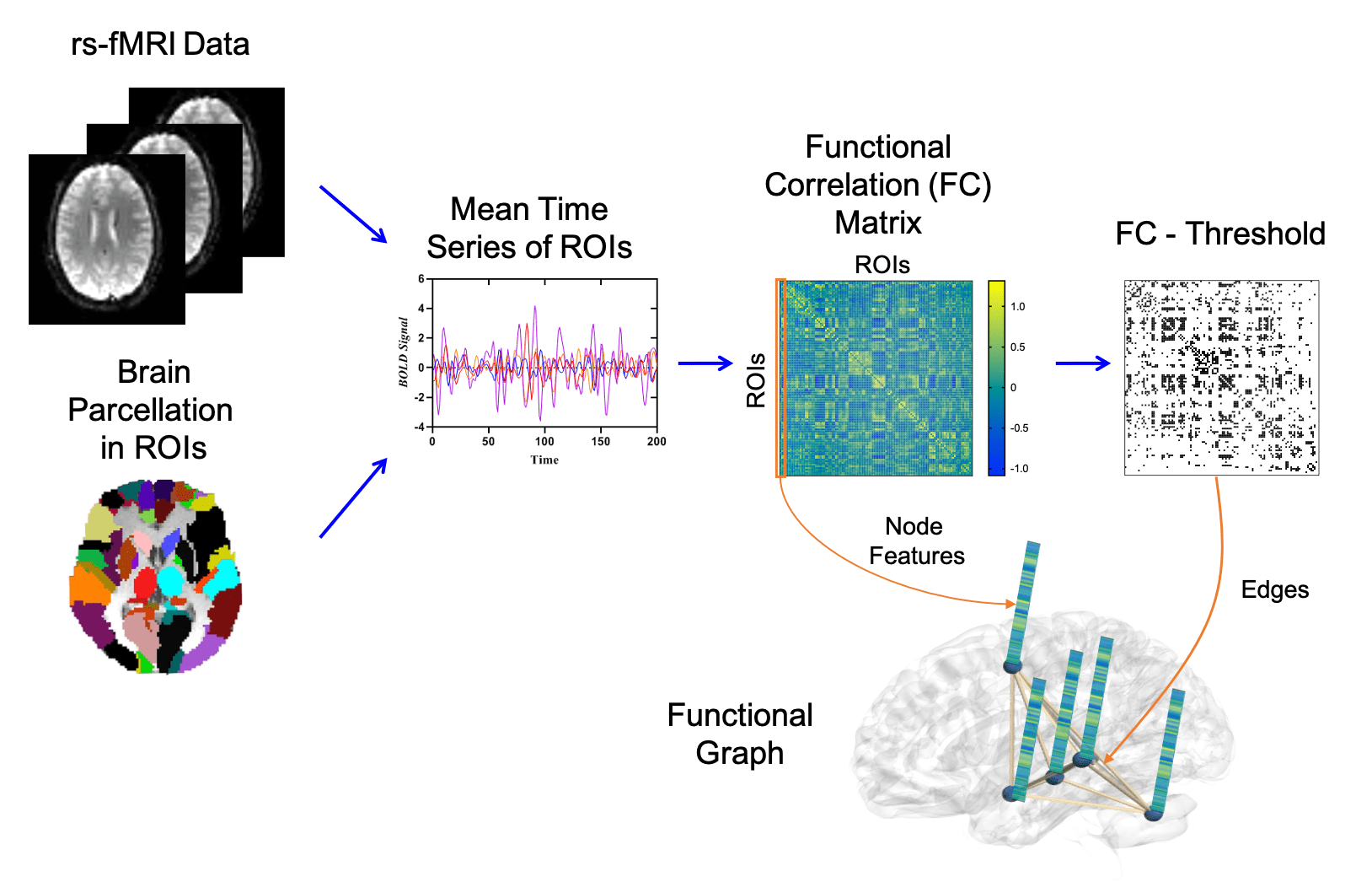}
\caption{Pipeline from rs-fMRI raw data to the functional graph. The resting-state fMRI raw data are preprocessed and then parcellated into regions of interest (ROIs) according to the automated anatomical labelling (AAL) atlas. The functional connectivity (FC) matrices are calculated using Pearson correlation between ROIs.  From the FC we construct the brain functional graph $\mathcal{G}=\left(A,X \right)$, where $A$ is the graph adjacency matrix characterizing the graph structure ($A\in\left\{ 0,1\right\}^{n\times n}$) and $X$ is node feature matrix. Specifically, $A$ is a binarized FC matrix, where only the top 20-percentile absolute values of the correlations of the matrix are transformed into ones, while the rest are transformed into zeros. For node feature $X$, $X_{k}$ for node $k$ can be defined as $X_{k}=\left [ \rho_{k1},\dots, \rho_{kn}\right ] ^{\text{T}}$, where $\rho_{kl}$ is the Pearson’s correlation coefficient for node $k$ and node $l$.}
\label{fig:graph}
\end{figure*}

Given a set of weighted brain networks $\left\{ \mathcal{G}^{1},\mathcal{G}^{2},...,\mathcal{G}^{N}\right\}$, the model outputs corresponding labels $\left\{ y^{1},y^{2},...,y^{N}\right\}$. We define the $i$-th brain network as $\mathcal{G}^{i}=\left(A^{i},X^{i} \right)$, where $A^{i}$ is the graph adjacency matrix characterizing the graph structure ($A^{i}\in\left\{ 0,1\right\}^{n\times n}$) and $X^{i}$ is node feature matrix constituted by weighted functional connectivity values ($X^{i}\in \mathbb{R} ^{n\times n}$). Specifically, $A^{i}$ is a binarized FC matrix, where only the top 20-percentile absolute values of the correlations of the matrix are transformed into ones, while the rest are transformed into zeros. For node feature $X$, $X^{i}_{k}$ for node $k$ can be defined as $X^{i}_{k}=\left [ \rho_{k1},\dots, \rho_{kn}\right ] ^{\text{T}}$ , where $\rho_{kl}$ is the Pearson’s correlation coefficient for node $k$ and node $l$. Fig.~\ref{fig:graph} shows the pipeline from rs-fMRI raw data to the brain functional graph. In brain network analysis, $N$ is the number of participants and $n$ is the number of regions of interest (ROIs). Note that, we only consider functional connectivity values as node features, which is common in brain network analysis~\cite{gallo2023functional}. 


\subsection{Overall framework of BrainIB}
\begin{figure*}[ht!]
\centering
\includegraphics[scale = 0.58]{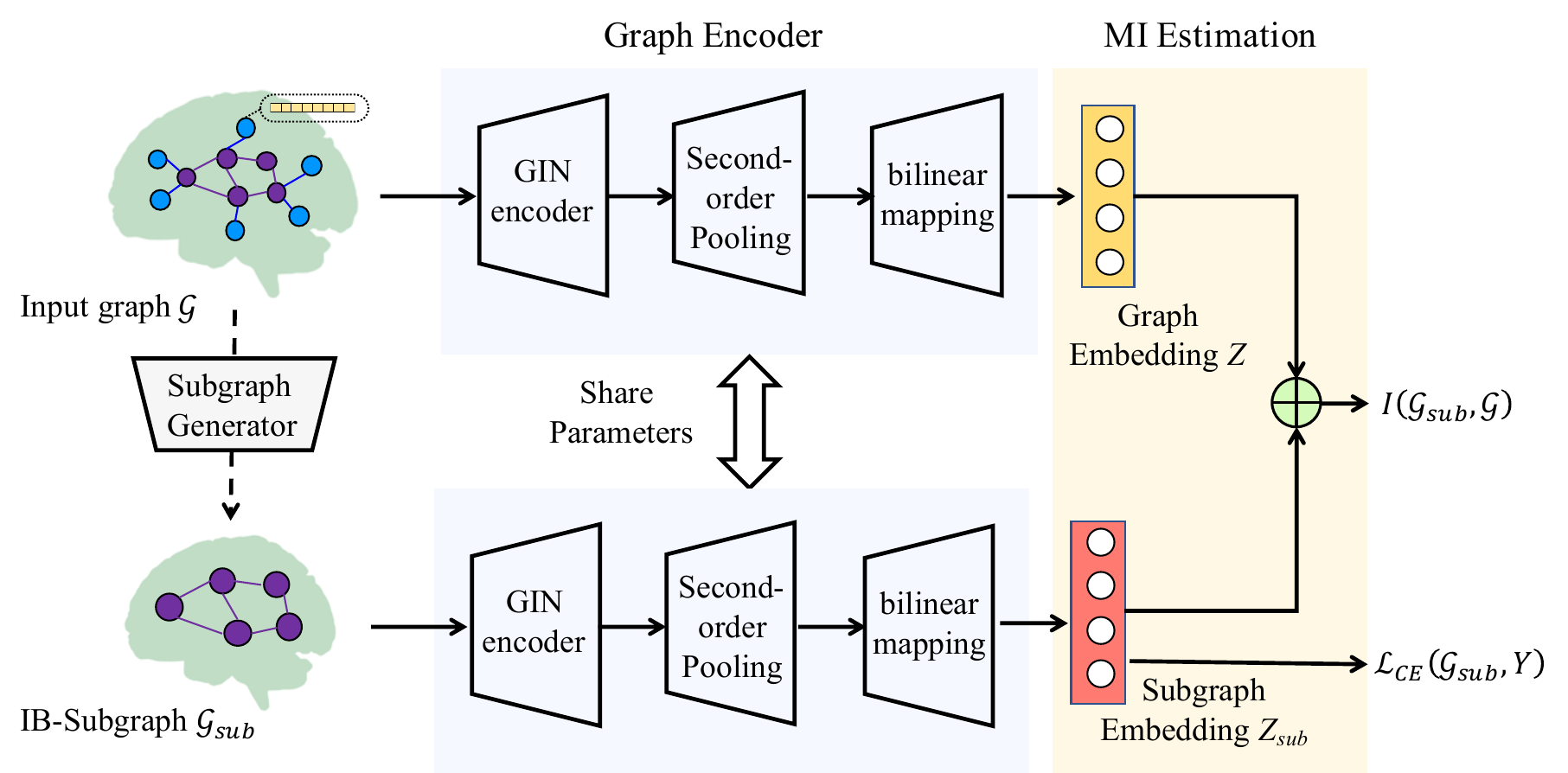}
\caption{Architecture of our proposed BrainIB. The framework mainly consists of three modules: subgraph generator module, graph encoder module and mutual information estimator module. Given an input graph $\mathcal{G}$ and corresponding graph label $Y$, the subgraph generator is used to sample subgraph $\mathcal{G}_{\text{sub}}$ from input graph $\mathcal{G}$. The graph encoder is used to learn graph embeddings $Z$ or $Z_{\text{sub}}$ from either $\mathcal{G}$ or $\mathcal{G}_{\text{sub}}$, where $Z_{\text{sub}}$ be used for graph classification by computing cross-entropy loss $\mathcal{L}_{\text{CE}}\left ( \mathcal{G}_{\text{sub}},Y \right )$. The mutual information estimation module evaluates the mutual information $I\left ( \mathcal{G}_{\text{sub}},\mathcal{G} \right )$ between $\mathcal{G}$ and $\mathcal{G}_{\text{sub}}$.}
\label{fig:brainIB_framework}
\end{figure*}

The flowchart of BrainIB is illustrated in Fig.~\ref{fig:brainIB_framework}. BrainIB consists of three modules: subgraph generator, graph encoder, and mutual information estimation module. The subgraph generator is used to sample subgraph $\mathcal{G}_{\text{sub}}$ from the original $\mathcal{G}$. The graph encoder is used to learn graph embeddings from either $\mathcal{G}$ or $\mathcal{G}_{\text{sub}}$.
The mutual information estimation module evaluates the mutual information between $\mathcal{G}$ or $\mathcal{G}_{\text{sub}}$.

\subsection{Subgraph Generator}
\begin{figure*}[ht!]
\centering
\includegraphics[scale = 0.6]{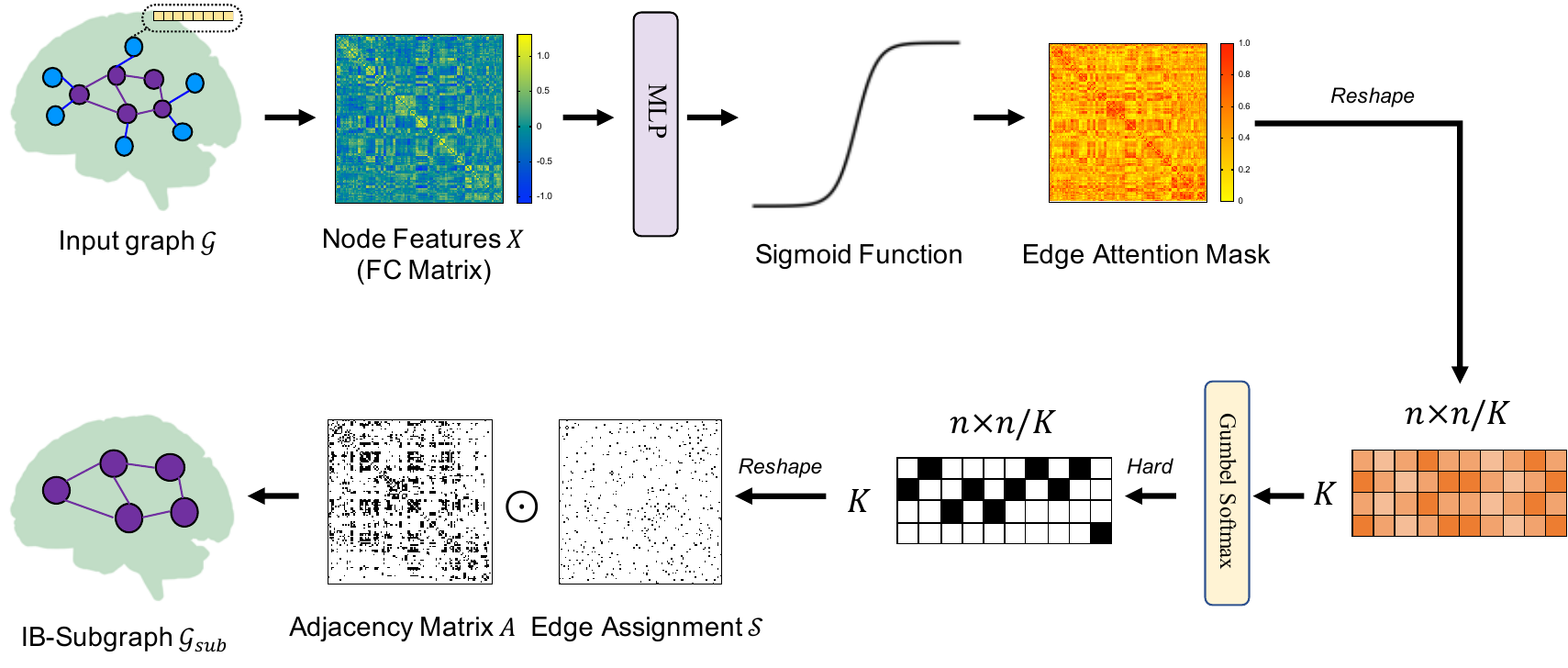}
\caption{Subgraph Generator. IB-Subgraph is generated from the input graph with the edge selection strategy. Given an input graph $\mathcal{G}=\left(A,X \right)$, $X$ is sent to a linear layer followed by a sigmoid function to obtain edge attention mask $P$ and ensure $p_{ij} \in\left [ 0,1 \right ]$, where $p_{ij}$ represents edge selection probability between node $i$ and node $j$. Subsequently, we binarize $P$ to obtain edge assignment $\mathcal{S}$. In order to make sure the gradient, w.r.t., $p_{ij}$ is computable, we leverage the gumbel-softmax reparameterization trick to update edge assignment $\mathcal{S}$. To ensure that a sufficient number of edges are retained, $P$ is reshaped to $K$-dimensional matrix and then is binarized to generate edge assignment $\mathcal{S}$ with Gumbel-Softmax approach. Next, $\mathcal{S}$ is transform to $n \times n$ matrix. Finally, the IB-subgraph $\mathcal{G}_{\text{sub}}$ can be extracted through $A_{\text{sub}}=A\odot \mathcal{S}$, where  $\odot$ is entry-wise product and $A$, $A_{\text{sub}}$  represent adjacency matrix of the input graph $\mathcal{G}$, IB-subgraph $\mathcal{G}_{\text{sub}}$, respectively.}
\label{fig:subgraph}
\end{figure*}

The procedure of subgraph generator module is shown in Fig.~\ref{fig:subgraph}. We generate IB-Subgraph from the input graph with edge assignment rather than node assignment. Here, edge assignments and node assignments indicate each edge and node is either in or out of the IB-subgraph, respectively. Given a graph $\mathcal{G}=\left(A,X \right)$, we calculate the probability of edges to determine the edge assignment $\mathcal{S}$ from node feature $X$. First, we learn edge attention mask $P$, where each entry represents edge selection probability. For simplicity, $X$ is directly sent to a Multi-layer Perceptron (MLP) followed by a sigmoid function to obtain edge attention mask $P\in \mathbb{R}^{n \times n}$ and ensure $p_{ij}\in\left [ 0,1 \right ]$, where $n$ is the number of node and $p_{ij}$ represents edge selection probability between node $i$ and node $j$.  Here, $p_{ij}$ is defines as:
\begin{equation}
p_{ij}=\sigma \left ( X_{i} W_{j}^{T}\right ) ,
\end{equation}
where $\sigma \left ( \cdot  \right )$ is sigmoid function, $X_{i}=\left [ \rho_{i1},\dots, \rho_{in}\right ]$ , $\rho_{ij}$ is the Pearson’s correlation coefficient for node $i$ and node $j$, and $W_{j}^{T}$ represents the parameters at $j$-th neuron for each layer of MLP. Here, the physical meaning is that the probability $p_{ij}$ on the existence of edge connecting node $i$ and node $j$ could be computed as the node feature in the $i$-th node through the $j$-th neuron of MLP.

Subsequently, we binarize $P$ to obtain edge assignment $\mathcal{S} \in \left \{ 0,1 \right \}^{n\times n} $. In order to make sure the gradient, w.r.t., $p_{ij}$ is computable, we leverage the gumbel-softmax reparameterization trick~\cite{maddison2017concrete} to update edge assignment $\mathcal{S}$. In general, Gumbel-Softmax approximates a continuous random variable represented as a one-hot vector. However, if we directly apply the Gumbel softmax function to $P$, we can only select one edge for continuous $n$ edges.

To ensure that a sufficient number of edges are retained, $P$ is reshaped to $K$-dimensional matrix and then is binarized to generate edge assignment $\mathcal{S}$ with Gumbel-Softmax approach. This way ensures that at least one edge is selected from $K$ consecutive edges. Here, $k$-th edge sample probability for the $K$-dimensional sample vector is defined as:
\begin{equation}
\hat{p}_{k}=\frac{\mbox{exp}\left ( \left ( \mbox{log}p_{k}+c_{k} \right )/\tau  \right )}{\sum_{i=1}^{K}\mbox{exp}\left ( \left ( \mbox{log}p_{i}+c_{i} \right )/\tau  \right ) },
\end{equation}
where $\tau$ is a temperature for the Concrete distribution,  $p$ is edge selection probability, $\hat{p}$ is sample  probability and $c_{k}$ is generated from a Gumbel(0, 1) distribution:
\begin{equation}
c_{k}=-\mbox{log}\left ( -\mbox{log} U_{k} \right ),U_{k}\sim \mbox{Uniform}(0,1).
\end{equation}

Thus, according to this procedure, $K$ is able to determine the size of IB-subgraph $\mathcal{G}_{\text{sub}}$ and we could preserve $n \times n/K$ edges. Next, $\mathcal{S}$ is transform to $n \times n$ matrix. Finally, the IB-subgraph $\mathcal{G}_{\text{sub}}$ can be extracted through $A_{\text{sub}}=A\odot \mathcal{S}$, where $\odot$ is entry-wise product and $A$, $A_{\text{sub}}$  represent adjacency matrix of the input graph $\mathcal{G}$, IB-subgraph $\mathcal{G}_{\text{sub}}$, respectively.





\subsection{Graph Encoder}
Graph encoder module consists of GIN encoder and the bilinear mapping second-order pooling~\cite{wang2020second}. After GIN encoder, we obtain node representations $H\in \mathbb{R}^{n\times d}$ from the original node feature matrix $X$. 
We then apply the bilinear mapping second-order pooling to generate vectorized graph embeddings from $H$. Compared to existing graph pooling methods, bilinear mapping second-order pooling is able to use information from all nodes, collect second-order statistics and reduce the excessive number of training parameters~\cite{wang2020second}. For clarity, we first provide the definition of the bilinear mapping second-order pooling.

\begin{definition}
Given $H =\left [ h_{1}, h_{2},..., h_{n} \right ]^{T} \in \mathbb{R}^{n \times d}$ , the second-order pooling ($\mbox{SOPOOL}$) is defined as:
\begin{equation}
\begin{aligned}
\mbox{SOPOOL}\left ( H \right )= H^{T}H\in\mathbb{R}^{d \times d}.
\end{aligned}
\end{equation}
\end{definition}

\begin{definition}
Given $H =\left [ h_{1}, h_{2},..., h_{n} \right ]^{T} \in \mathbb{R}^{n \times d}$ and $W \in \mathbb{R}^{d \times d'}$ a trainable matrix representing a linear mapping from $\mathbb{R}^d$ to $\mathbb{R}^{d'}$, the bilinear mapping second-order pooling ($\mbox{SOPOOL}_{\mbox{bimap}}$) is formulated as:
\begin{equation}
\begin{aligned}
\mbox{SOPOOL}_{\mbox{bimap}}\left ( H \right )&= \mbox{SOPOOL}\left ( HW \right )
\\ &= W^{T}H^{T}HW\in\mathbb{R}^{d' \times d'}.
\end{aligned}
\end{equation}
\end{definition}

We directly apply $\mbox{SOPOOL}_{\mbox{bimap}}$ on $H=\mbox{GIN}\left (A,X \right)$ and flatten the output matrix into a $d'^{2}$-dimensional graph embedding $h_{g}$:
\begin{equation}
h_{g}=\mbox{FLATTEN}\left ( \mbox{SOPOOL}_{\mbox{bimap}}\left ( \mbox{GIN}\left ( A,X \right ) \right ) \right )\in \mathbb{R}^{d'^{2}}.
\end{equation}


\subsection{Mutual Information Estimation}

The graph IB objective includes two mutual information terms:
\begin{equation}
\min_{\mathcal{G}_{\text{sub}}} I\left ( \mathcal{G}_{\text{sub}},\mathcal{G} \right ) - \beta I\left ( \mathcal{G}_{\text{sub}},Y \right ).
\end{equation}

Minimizing $-I\left ( \mathcal{G}_{\text{sub}},Y \right )$ (\emph{a.k.a.}, maximizing $I\left ( \mathcal{G}_{\text{sub}},Y \right )$) encourages $\mathcal{G}_{\text{sub}}$ is most predictable to graph label $Y$. Mathematically, we have:
\begin{equation}\label{eq:variational}
\begin{aligned}
-I\left ( \mathcal{G}_{\text{sub}},Y \right )&\leq \mathbb{E}_{Y,\mathcal{G}_{\text{sub}}} -\mbox{log} q_{\theta }\left ( Y|\mathcal{G}_{\text{sub}} \right )\\ 
&:=\mathcal{L}_{\text{CE}}\left ( \mathcal{G}_{\text{sub}},Y \right ),
\end{aligned}
\end{equation}
where $q_{\theta }\left ( Y|\mathcal{G}_{\text{sub}} \right )$ is the variational approximation to the true mapping $p\left ( Y|\mathcal{G}_{\text{sub}} \right )$ from $\mathcal{G}_{\text{sub}}$ to $Y$. Eq.~(\ref{eq:variational}) indicates that $\min -I\left ( \mathcal{G}_{\text{sub}},Y \right )$ approximately equals to minimizing the cross-entropy loss $\mathcal{L}_{\text{CE}}$.


As for the mutual information term $I\left ( \mathcal{G}_{\text{sub}},\mathcal{G} \right )$, we first obtain (vectorized) graph embeddings $Z_{\text{sub}}$ and $Z$ from respectively $\mathcal{G}_{\text{sub}}$ and $\mathcal{G}$ by the graph encoder. According to the sufficient encoder assumption~\cite{tian2020makes} that the information of $Z$ is lossless in the encoding process, we approximate $I\left ( \mathcal{G}_{\text{sub}},\mathcal{G} \right )$ with $I\left ( Z_{\text{sub}},Z \right ) $. Different from SIB that uses MINE which requires an additional neural network, we directly estimate $I\left ( Z_{\text{sub}},Z \right)$ with the recently proposed matrix-based R{\'e}nyi's $\alpha $-order mutual information~\cite{giraldo2014measures,yu2019multivariate}, which is mathematically well-defined and computationally efficient. 

Specifically, given a mini-batch of samples of size $N$, we obtain both $\{Z^i\}_{i=1}^N$ and $\{Z_{\text{sub}}^i\}_{i=1}^N$, in which $Z^i$ and $Z_{\text{sub}}^i$ refer to respectively the graph embeddings of the $i$-th graph and the $i$-th subgraph (in a mini-batch). According to ~\cite{giraldo2014measures}, 
we can evaluate the entropy of graph embeddings using the eigenspectrum of the (normalized) Gram matrix  $D$ obtained from $\{Z^i\}_{i=1}^N$ as:
\begin{equation}\label{eq:entropy}
\begin{aligned}
H_{\alpha }\left ( Z \right )&=\frac{1}{1-\alpha }\log_{2}\left ( \tr \left ( D^{\alpha } \right ) \right ) \\ 
&=\frac{1}{1-\alpha }\log_{2}\left (\sum_{i=1}^{N}\lambda_{i}\left ( D \right )^{\alpha }  \right ),
\end{aligned}
\end{equation}
where $\tr$ denotes trace of a matrix, $D=K/\tr (K)$, and $K=\kappa \left (Z^{i}, Z^{j} \right )$ is the Gram matrix obtained from $\{Z^i\}_{i=1}^N$ with a positive definite kernel $\kappa$ on all pairs of exemplars. $\lambda_i$ denotes the $i$-th eigenvalue of  $D$. In the limit case of $\alpha \to 1$, Eq.~(\ref{eq:entropy}) reduced to an entropy-like measure that resembles the Shannon entropy of $H(Z)$.

Similarly, we can evaluate the entropy of subgraph embeddings from $\{Z_{\text{sub}}^i\}_{i=1}^N$ by:
\begin{equation}
\begin{aligned}
H_{\alpha }\left ( Z_{\text{sub}} \right )&=\frac{1}{1-\alpha }\log_{2}\left ( \tr \left ( D_{\text{sub}}^{\alpha } \right ) \right ) \\ 
&=\frac{1}{1-\alpha }\log_{2}\left (\sum_{i=1}^{N}\lambda_{i}\left ( D_{\text{sub}} \right )^{\alpha }  \right ),
\end{aligned}
\end{equation}
where $D_{\text{sub}}$ is the trace normalized Gram matrix evaluated on $\{Z_{\text{sub}}^i\}_{i=1}^N$ also with the kernel function $\kappa$.

The joint entropy between $Z$ and $Z_{\text{sub}}$ can be evaluated as:
\begin{equation}\label{eq:mutual_information}
H_{\alpha }\left ( Z,Z_{\text{sub}} \right )=H_{\alpha }\left ( \frac{D\circ D_{\text{sub}}}{\tr \left ( D\circ D_{\text{sub}} \right )} \right ),
\end{equation}
in which $D\circ D_{\text{sub}}$ denotes the Hadamard product between the $D$ and $D_{\text{sub}}$.

According to Eqs.~(\ref{eq:entropy})-(\ref{eq:mutual_information}), the matrix-based Rényi’s $\alpha$-order mutual information $I\left ( Z_{sub},Z \right )$ in analogy of Shannon’s mutual information is defined as~\cite{yu2019multivariate}:
\begin{equation}
I\left ( Z_{\text{sub}},Z \right )=H_{\alpha }\left ( Z_{\text{sub}} \right )+ H_{\alpha }\left ( Z \right )-H_{\alpha }\left ( Z_{\text{sub}}, Z \right ).
\end{equation}

In this work, we use the radial basis function (RBF) kernel $\kappa$ to obtain  $D$ and $D_{\text{sub}}$:
\begin{equation}
\kappa \left ( z^{i},z^{j} \right ) = \mbox{exp}\left ( -\frac{\left\|z^{i}-z^{j} \right\|^{2}}{2\sigma ^{2}} \right ),
\end{equation}
and for the kernel width $\sigma$, we estimate the $k$ ($k$ = 10) nearest distances of each sample and obtain the mean. We set the $\sigma$ with the average of mean values for all samples. We also fix $\alpha=1.01$ to approximate Shannon mutual information.

The final objective of BrainIB can be formulated as:
\begin{equation}\label{eq:brainIB_objective}
\mathcal{L}=\mathcal{L}_{\text{CE}}\left ( \mathcal{G}_{\text{sub}},Y \right )+\beta I\left ( \mathcal{G}_{\text{sub}},\mathcal{G} \right )
\end{equation}
in which  $\beta$ are the hyper-parameters.

\section{Experiments}
In order to demonstrate the effectiveness and superiority of BrainIB, we compare the performance of BrainIB against $10$ popular brain network classification models on three psychiatric datasets, namely ABIDE,  REST-meta-MDD and SRPBS. 
 The selected competitors include: a benchmark SVM model~\cite{2015Functional}; three baselines including GCN~\cite{kipf2016semi}, GAT~\cite{velivckovic2018graph} and GIN~\cite{xu2018powerful}; three popular self-explainable graph neural networks (GNNs) including SIB~\cite{yu2021recognizing}, DIR-GNN~\cite{wudiscovering} and ProtGNN~\cite{zhang2022protgnn} that use GCN as the backbone network; and three state-of-the-art (SOTA) GNNs designed for brain networks including BrainGNN~\cite{li2021braingnn}, IBGNN~\cite{cui2022interpretable} and CI-GNN~\cite{zheng2024ci}.
We perform both $10$-fold and leave-one-site-out cross validations to assess the generalization capacity of BrainIB to unseen data. We also evaluate the interpretability of BrainIB with respect to clinical and neuroimaging findings.

\subsection{Data sets and Data Preprocessing}
 We use ABIDE, REST-meta-MDD and SRPBS datasets. Their demographic information is summarized in Table~\ref{tab:Demographic}. 

\begin{table}[ht!]
\caption{Demographic and clinical characteristics.}
\renewcommand\arraystretch{1}
\resizebox{\linewidth}{!}{\begin{tabular}{@{}c|cccccc@{}}
\toprule[1.5pt]
\multirow{2}{*}{\textbf{Characteristic}} & \multicolumn{2}{c}{\textbf{ABIDE}} & \multicolumn{2}{c}{\textbf{Rest-meta-MDD}} & \multicolumn{2}{c}{\textbf{SRPBS}}    \\ \cmidrule(l){2-7} 
                                & ASD         & TD          & MDD             & HC              & Schizophrenia & HC           \\ \midrule
Sample Size                     & 528         & 571         & 828             & 776             & 92            & 92           \\
Age                             & 17.0 ± 8.4  & 17.1 ± 7.7 & 34.3 ± 11.5     & 34.4 ± 13.0     & 39.6 ± 10.4   & 38.0 ± 12.4 \\
Gender (M/F)                    & 464/64      & 471/100      & 301/527         & 318/458         & 47/45         & 60/32        \\ \bottomrule[1.5pt]
\end{tabular}}\label{tab:clinical}
\begin{tablenotes}
 \footnotesize
 \item  ``-'' denotes the missing values.
\end{tablenotes}
\label{tab:Demographic}
\end{table}

The Autism Brain Imaging Data Exchange I (ABIDE-I)~\cite{di2014autism} dataset is a grassroots consortium aggregating and openly sharing more than $1,000$ existing resting-state fMRI data\footnote{\url{http://fcon_1000.projects.nitrc.org/indi/abide/}}. In this study, a total of $528$ patients with ASD and $571$ typically developed (TD) individuals were provided by ABIDE. Our further analysis was based on resting-state raw fMRI data and these data were preprocessed using the statistical parametric mapping (SPM) software\footnote{\url{https://www.fil.ion.ucl.ac.uk/spm/software/spm12/}}. First, the resting-state fMRI images are corrected for slice timing, compensating the differences in acquisition time between slices. Then, realignment is performed to correct for head motion between fMRI images at different time point by translation and rotation. The deformation parameters from the fMRI images to the MNI (Montreal Neurological Institute) template are then used to normalize the resting-state fMRI images into a standard space. Additionally, a Gaussian filter with a half maximum width of 6 mm is used to smooth the functional images. The resulting fMRI images are temporally filtered with a band-pass filter ($0.01$–$0.08$ Hz). Finally, we regress out the effects of head motion, white matter and cerebrospinal fluid signals (CSF).

The REST-meta-MDD is the largest MDD R-fMRI database to date~\cite{yan2019reduced}. It contains fMRI images of $2,428$ participants ($1,300$ patients with MDD and $1,128$ HCs) collected from twenty-five research groups from $17$ hospitals in China. In the current study, we select $1,604$ participants ($848$ MDDs and $794$ HCs) according to exclusion criteria from a previous study including incomplete information, bad spatial normalization, bad coverage, excessive head motion, and sites with fewer than $10$ subjects in either group and incomplete time series data. Our further analysis was based on preprocessed data previously made available by the REST-meta-MDD. The preprocessing pipeline includes discarding the initial $10$ volumes, slice-timing correction, head motion correction, space normalization, temporal bandpass filtering (0.01-0.1 HZ) and removing the effects of head motion, global signal, white matter and cerebrospinal fluid signals, as well as linear trends.

The SRPBS provides open access to over 1000 resting-state fMRI data from patients with psychiatric disorders and healthy controls in Japan\footnote{\url{https://bicr-resource.atr.jp/srpbsfc/}}. For our study, we chose a sample of 184 participants, consisting of 92 patients with schizophrenia and 92 healthy controls. The preprocessing steps used are the same as those applied to the ABIDE dataset.
 
After preprocessing, mean time series of each participant is extracted from each region of interests (ROIs) by the automated anatomical labelling (AAL) template, which consists of $90$ cerebrum regions and $26$ cerebellum regions. In addition, we compute functional connectivity (Fisher’s r-to-z transformed Pearson’s correlation) between all ROI pairs to generate a $116\times 116$ symmetric matrix (brain network).

\subsection{BrainIB Configurations and Hyperparameter Setup}
 We train and test the BrainIB with PyTorch 1.12.1\footnote{https://github.com/SJYuCNEL/brain-and-Information-Bottleneck}. During the training, the number of epoch is set to 300 and dropout ratio is set to 0.5. Table~\ref{tab:range} demonstrates the range of hyper-parameters that are examined and the final specification of all hyper-parameters are applied to generate the reported results. The hyper-parameters are set through a grid search or based on the recommended settings of related work.

\begin{table}[ht]
\centering
\caption{Range of hyper-parameters and final specification for BrainIB.}
\footnotesize
\renewcommand\arraystretch{1}
\begin{tabular}{@{}ccc@{}}
\toprule
Hyper-parameter               & Range Examined                     & Final Specification     \\ \midrule
\textbf{Graph Encoder}\\\midrule
\#GNN Layers                 & {[}2,3,4,5,6{]}                      & 5                       \\
\#MLP Layers                  & {[}2,3,4{]}                      & 2                       \\
\#Hidden Dimensions           & {[}64,128,256,512{]}               & 128            \\\midrule
\textbf{Subgraph Generator}\\ \midrule
\#MLP Layers                  & {[}2,3,4{]}                      & 2                       \\
\#Hidden Dimensions           & {[}16,32,64{]}               & 16\\   \midrule
Learning Rate                 & {[}1e-2,1e-3,1e-4{]}               & 1e-3                    \\
Batch Size                    & {[}32,64,128{]}                    & 32                      \\
Weight Decay                  & {[}1e-3,1e-4{]}                    & 1e-4                  \\ 
$K$                             & {[}2,4,8{]}                    & 2                 \\ 
\bottomrule
\end{tabular}
\label{tab:range}
\end{table}

For the baselines, we train each model with 300 epochs. For GIN,  Graph Attention Network (GAT) and SIB, we use authors' recommended hyperparameters to train the models. Additionally, we tune the weight $\beta$ of the mutual information term $I(\mathcal{G}_{\text{sub}};\mathcal{G})$ in the SIB objective in the range $ \left\{0.0001,0.1\right\}$ and select $\beta=0.001$ finally.


\subsection{Hyperparameter Sensitivity Analysis and Ablation Study}
\begin{figure}[ht!]
  \centering
  \subfloat[\footnotesize{ABIDE}]{\includegraphics[scale=0.4]{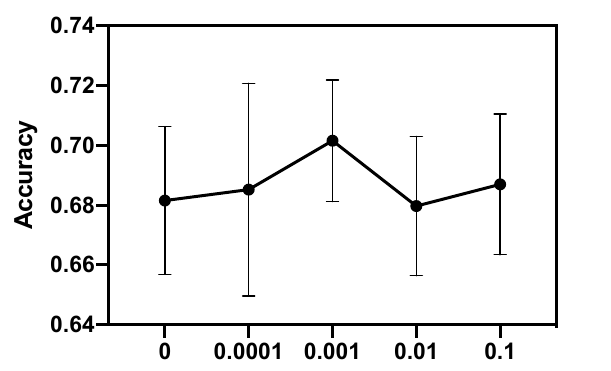}\label{fig:sensitivityA}}
  \hfil
  \subfloat[\footnotesize{REST-meta-MDD}]{\includegraphics[scale=0.4]{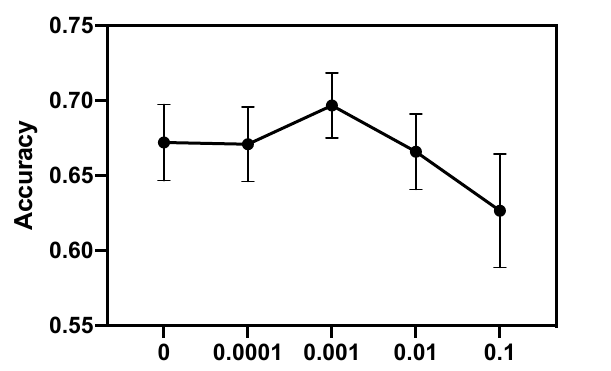}\label{fig:sensitivityR}}
  \hfil
  \caption{The influence of hyperparameter $\beta$ on performance using the ABIDE and Rest-meta-MDD datasets. }
  \label{fig:Sensitivity}
\end{figure}

To analyze how the hyperparameter $\beta $ affect the model performance, we further tune $\beta  \in \left \{ 0,0.0001,0.001,0.01,0.1 \right \}$ in the loss function using the tenfold cross validation on the ABIDE and Rest-meta-MDD datasets. Fig.~\ref{fig:Sensitivity} demonstrates the mean and standard deviation of accuracy for different $\beta$ across 10 folds. As shown in Fig.~\ref{fig:Sensitivity}, we observe that BrainIB achieves the highest classification accuracy on both brain disease datasets, when $\beta$ is set as 0.001.

\graphicspath{{./figs/}}

\begin{figure}[ht!]
  \centering
  \subfloat[\footnotesize{ABIDE}]{\includegraphics[scale=0.52]{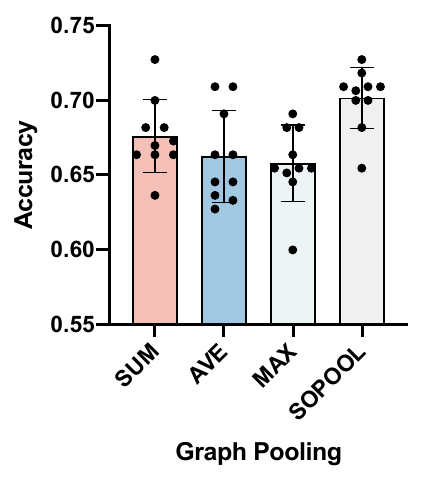}\label{fig:poolingA}}
  \hfil
  \subfloat[\footnotesize{REST-meta-MDD}]{\includegraphics[scale=0.52]{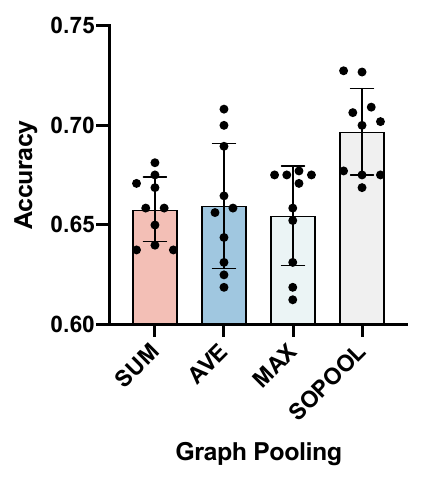}\label{fig:poolingR}}
  \hfil
  \caption{The influence of different graph pooling methods on performance using the ABIDE and Rest-meta-MDD datasets. SUM, AVE, MAX and SOPOOL represent sum-pooling, average-pooling, max-pooling and bilinear mapping second-order pooling, respectively.}
  \label{fig:pooling}
\end{figure}

Furthermore, we perform ablation studies to show whether bilinear mapping second-order pooling is able to improve the model performance. Specifically, we compare bilinear mapping second-order pooling with other graph pooling methods including sum-pooling, average-pooling and  max-pooling by fixing other BrainIB configurations. Fig.~\ref{fig:pooling} provides the comparison results. We observe that bilinear mapping second-order pooling achieves better classification performance than sum-pooling, average-pooling and  max-pooling.

\graphicspath{{./figs/}}

\subsection{Computational Complexity of BrainIB}

We analyze the computational and memory cost of BrainIB. Let $d$ be the dimension of features and $n$ be the number of samples (i.e., mini-batch size in our context). The main time complexity of BrainIB comes from the estimation of  $I\left (\mathcal{G}_{\text{sub}},\mathcal{G} \right )$. BrainIB takes time for the $\mathcal{O}\left ( n^{3} \right ) $ eigenvalue decomposition of a $n\times n$ Gram matrix. Thus, the computational cost of BrainIB is $\mathcal{O}\left ( n^{3} \right ) $. As for the memory cost, BrainIB needs to reserve Gram matrices of size $n\times n$, resulting in $\mathcal{O}\left ( n^{2} \right ) $. Furthermore, we compare the computational complexity of BrainIB with DIR-GNN, CI-GNN, SIB and report the time required per epoch on three psychiatric datasets for each model in Table~\ref{tab:computation}. The main computational complexity of CI-GNN comes from the estimation of conditional mutual information, resulting in $\mathcal{O}\left ( n^{3} \right ) $. However, the computation of multiple matrix-based R{\'e}nyi's $\alpha$-order entropy functional and the use of a two-step training strategy in CI-GNN result in a longer running time compared to BrainIB. For DIR-GNN, the main computational complexity of DIR-GNN comes from rationale generator and distribution intervener, which use causal strategy to find invariant rationale edges. This method may take $\mathcal{O}\left ( m  n \right ) $ time, where $m$ is the number of edges. As $m$ and $n^{2}$ are of the same order of magnitude, the training time for BrainIB and DIR-GNN is similar. Moreover, BrainIB has a lower computational cost compared with SIB. This is because SIB utilizes additional neural networks for  the estimation of $I\left (\mathcal{G}_{\text{sub}},\mathcal{G} \right )$ and employs a bilevel optimization scheme. These results indicate that BrainIB is acceptable for real-world viability.


\begin{table}[ht!]
\centering
\caption{ Training time per epoch (s) of different methods on three psychiatric datasets.}
\renewcommand\arraystretch{1}
\begin{tabular}{ccccc}
\hline
\textbf{Datasets}      & \textbf{DIR-GNN} & \textbf{CI-GNN} & \textbf{SIB}& \textbf{BrainIB} \\ \hline
\textbf{ABIDE}         & 3.94             & 52.33       & 28.39    & \textbf{1.84}    \\
\textbf{REST-meta-MDD} & 3.13             & 36.15      & 28.42     & \textbf{1.75}    \\
\textbf{SRPBS}         & 2.34             & 7.02       & 28.26     & \textbf{1.73}    \\ \hline
\end{tabular}
\label{tab:computation}
\end{table}

\subsection{Stable Training Discussion}
\label{stable}
\begin{figure}[ht!]
  \centering
  \subfloat[\footnotesize{SIB in ABIDE}]{\includegraphics[scale=0.37]{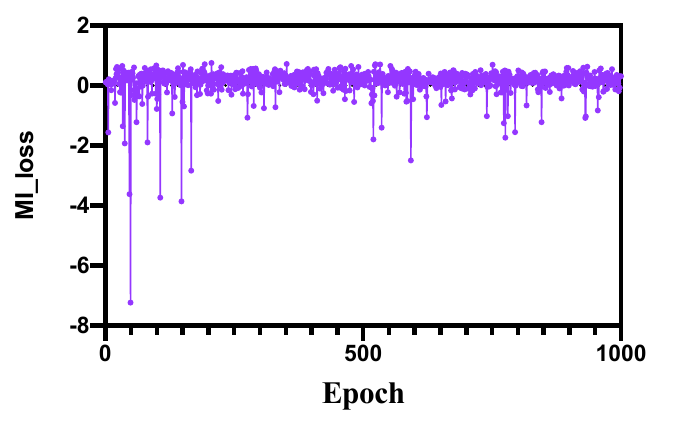}\label{fig:SIBA}}
  \hfil
  \subfloat[\footnotesize{SIB in REST-meta-MDD}]{\includegraphics[scale=0.37]{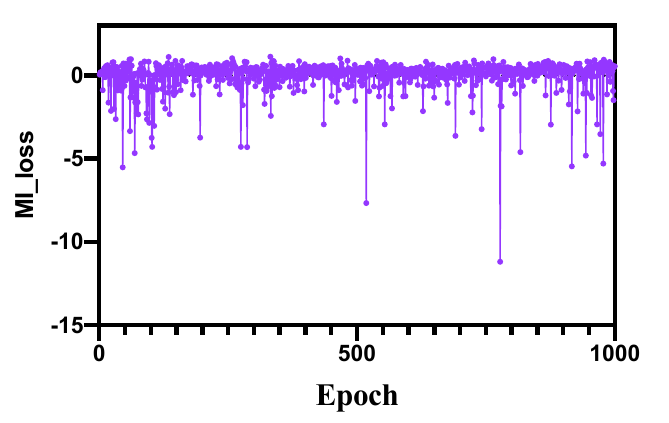}\label{fig:SIBR}}
  \hfil
  \subfloat[\footnotesize{BrainIB in ABIDE}]{\includegraphics[scale=0.37]{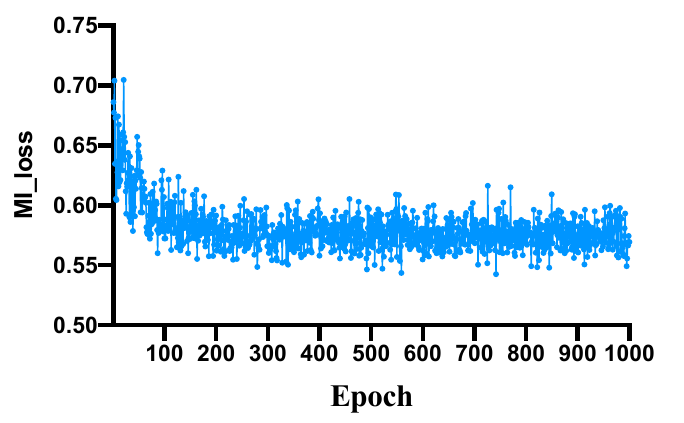}\label{fig:BrainIBA}}
  \hfil
  \subfloat[\footnotesize{BrainIB in Rest-meta-MDD}]{\includegraphics[scale=0.37]{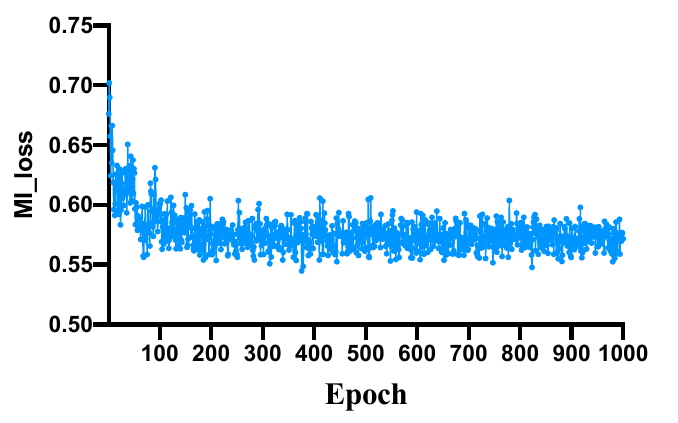}\label{fig:BrainIBR}}
  \hfil
    \subfloat[\footnotesize{VGIB in ABIDE}]{\includegraphics[scale=0.37]{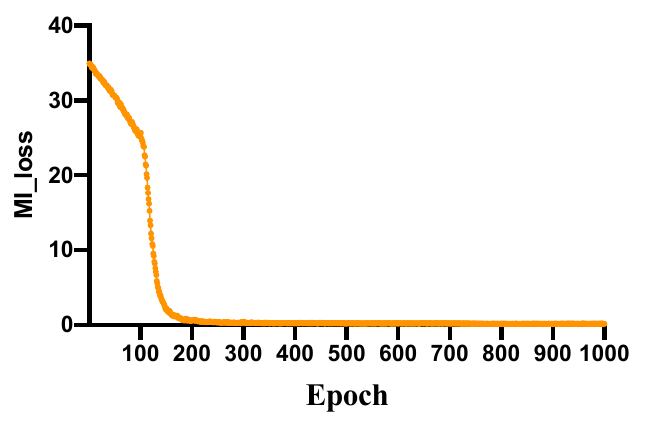}\label{fig:VGIBA}}
  \hfil
    \subfloat[\footnotesize{VGIB in Rest-meta-MDD}]{\includegraphics[scale=0.37]{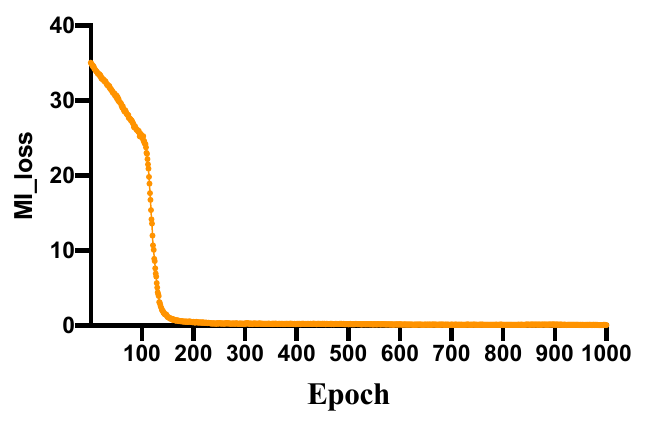}\label{fig:VGIBR}}
  \hfil
  \caption{Training dynamics of $I(\mathcal{G};\mathcal{G}_{\text{sub}})$ (MI\_loss) in SIB, BrainIB and VGIB over 1000 epochs on REST-meta-MDD and ABIDE.  $I(\mathcal{G};\mathcal{G}_{\text{sub}})$ is the mutual information between subgraph and input graph. The training process of BrainIB is stable, while SIB suffers from an unstable training process and inaccurate estimation of mutual information between subgraph and input graph. In addition, $I(\mathcal{G};\mathcal{G}_{\text{sub}})$ converges faster in BrainIB, compared with VGIB. }
  \label{fig:stable}
\end{figure}

Although SIB completes theoretical process of subgraph information bottleneck (see Eq.~(\ref{eq:GIB_Lagrangian})), its optimization process is unstable, especially mutual information estimation $I\left ( \mathcal{G}_{\text{sub}},\mathcal{G} \right )$, which is demonstrated in Figs.~\ref{fig:stable} (a) and \ref{fig:stable} (b). This phenomenon is caused by the use of the mutual information neural estimator (MINE)~\cite{belghazi2018mutual}, which needs to train auxiliary neural network $f$. To calculate $I\left ( \mathcal{G}_{\text{sub}},\mathcal{G} \right )$, graph embeddings $Z_{\text{sub}}$ and $Z$ are obtained from respectively $\mathcal{G}_{\text{sub}}$ and $\mathcal{G}$ by the graph encoder. According to the sufficient encoder assumption~\cite{tian2020makes} that the information of $Z$ is lossless in the encoding process, $I\left ( \mathcal{G}_{\text{sub}},\mathcal{G} \right )$ is approximated by $I\left ( Z_{\text{sub}},Z \right ) $. Thus, it is defined as:

\begin{equation}
\begin{aligned}
I\left ( \mathcal{G}, \mathcal{G}_{\text{sub}} \right )& \approx I\left ( \mathcal{Z}, \mathcal{Z}_{\text{sub}} \right ) \\&=\sup_{f\left ( \mathcal{Z}, \mathcal{Z}_{\text{sub}} \right ) }\mathbb{E}_{\mathcal{Z}, \mathcal{Z}_{\text{sub}} \in p\left ( \mathcal{Z}, \mathcal{Z}_{\text{sub}} \right ) } f\left ( \mathcal{Z}, \mathcal{Z}_{\text{sub}} \right )\\ &-\log \mathbb{E}_{\mathcal{Z} \in p\left ( \mathcal{Z} \right ) , \mathcal{Z}_{\text{sub}} \in p\left ( \mathcal{Z}_{\text{sub}} \right )} e^{ f\left ( \mathcal{Z}, \mathcal{Z}_{\text{sub}} \right )}.
\end{aligned}
\end{equation}

Due to the application of $f$, estimation of $I\left ( \mathcal{G}, \mathcal{G}_{\text{sub}} \right )$ is unstable during the training or even lead to negative values (see Figs.~\ref{fig:stable} (a) and \ref{fig:stable} (b)). To address this issue, we use matrix-based R{\'e}nyi's $\alpha$-order mutual information to estimate $I\left ( \mathcal{G}, \mathcal{G}_{\text{sub}} \right )$. To calculate $I\left ( \mathcal{G}_{\text{sub}},\mathcal{G} \right )$, we first obtain (vectorized) graph embeddings $Z_{\text{sub}}$ and $Z$ from respectively $\mathcal{G}_{\text{sub}}$ and $\mathcal{G}$ by the graph encoder. Then we approximate $I\left ( \mathcal{G}_{\text{sub}},\mathcal{G} \right )$ with $I\left ( Z_{\text{sub}},Z \right ) $. Because the R{\'e}nyi's $\alpha$-order entropy is defined in terms of the normalized eigenspectrum of the Gram matrix of the data projected to a reproducing kernel Hilbert space (RKHS)~\cite{giraldo2014measures,yu2019multivariate}, we could directly estimate $H_{\alpha}\left ( Z  \right )$ , $H_{\alpha}\left ( Z_{\text{sub}}  \right )$ and $H_{\alpha}\left ( Z, Z_{\text{sub}}  \right )$ from data without discrete probability density function (PDF) estimation and any auxiliary neural network. According to Shannon’s chain rule, $I\left ( Z, Z_{\text{sub}} \right )$ can be decomposed as:

\begin{equation}
I\left ( Z_{\text{sub}},Z \right )=H_{\alpha }\left ( Z_{\text{sub}} \right )+ H_{\alpha }\left ( Z \right )-H_{\alpha }\left ( Z_{\text{sub}}, Z \right ).
\end{equation}
Thus, $I\left ( \mathcal{G}, \mathcal{G}_{\text{sub}} \right )$ can be easily computed and is also differentiable which suits well for deep learning applications. In Figs.~\ref{fig:stable} (a-d), matrix-based R{\'e}nyi's $\alpha$-order mutual information significantly stabilizes the training compared with MINE.

Recently, Yu et al. also propose Variational Graph Information Bottleneck (VGIB)~\cite{yu2022improving} to address unstable training issues of SIB. VGIB employs a noise injection method to achieve a tractable variational upper bound of VGIB objective and effective optimization of the gradient-based method. However, VGIB uses the variational upper bound of $I\left ( G, G_{N} \right )$ to approximate $I\left ( \mathcal{G}, \mathcal{G}_{\text{sub}} \right )$, while BrainIB strictly calculates mutual information between $\mathcal{G}$ and $\mathcal{G}_{\text{sub}}$ , where $G_{N}$  is perturbed graph computed by the noise injection method. In addition, we compare the stable training of BrainIB and VGIB as shown in Fig.~\ref{fig:stable}. As can be seen, $I\left ( \mathcal{G}, \mathcal{G}_{\text{sub}} \right )$ in BrainIB (100 epoches) converges faster than VGIB (200 epoches). In addition, the decrease in $I\left ( \mathcal{G}, \mathcal{G}_{\text{sub}} \right )$ in brainIB (approximately 0.15) is smaller than VGIB (approximately 35). More importantly, although the approximation of $I\left ( \mathcal{G}, \mathcal{G}_{\text{sub}} \right )$ by VGIB significantly stablize the training, the final results may lack physical significance. As shown in Figs.~\ref{fig:stable} (e) and \ref{fig:stable} (f), the MI loss converges to almost zero after VGIB optimization. Such a result is highly unrealistic since $I\left ( \mathcal{G}, \mathcal{G}_{\text{sub}} \right )$ value of zero in the information bottleneck framework implies that $\mathcal{G}_{\text{sub}}$ does not receive any meaningful information from $\mathcal{G}$.

\graphicspath{{./figs/}}

\section{Results}
\subsection{Generalization Performance}

\subsubsection{Tenfold Cross Validation}
 Here, we implement tenfold cross validation to assess performance in three datasets including ABIDE, Rest-meta-MDD and SRPBS and accuracy is used as evaluating indicator. Table \ref{tab:ten-fold} shows the mean and standard deviation of accuracy for different models across 10 folds. 

As can be seen, BrainIB yields significant and consistent improvements over all SOTA baselines in both datasets. For MDD dataset, BrainIB outperforms traditional shallow model (SVM) with nearly 15.5\% absolute improvements. In addition, compared with other SOTA GNN models such as GIN, BrainIB achieves more than 5\% absolute improvements. For ASD dataset, BrainIB achieves the highest accuracy of 70.2\% compared with all SOTA shallow and deep baselines.  In addition, compared with recent GNNs designed for brain networks such as BrainGNN, BrainIB achieves more than 9\% absolute improvements on REST-meta-MDD, indicating the efficacy of our brain network-oriented design. More importantly, the superiority of BrainIB against SIB suggests the effectiveness of our improvement over SIB for brain networks analysis.


\begin{table}[ht]
\caption{Tenfold cross validation performance of different models on REST-meta-MDD, ABIDE and SRPBS. The highest performance is highlighted with boldface. All the performance of
methods are reported under their best settings.}
\centering
\renewcommand\arraystretch{1}
\resizebox{\linewidth}{!}{\begin{tabular}{@{}cccc@{}}
\toprule
\textbf{Methods}        & \textbf{REST-meta-MDD} & \textbf{ABIDE}         & \textbf{SRPBS}         \\ \midrule
SVM                     & 0.545 ± 0.001          & 0.550 ± 0.003          & 0.813 ± 0.003          \\
GCN                     & 0.608 ± 0.013          & 0.644 ± 0.038          & 0.827 ± 0.086          \\
GAT                     & 0.647 ± 0.017          & 0.679 ± 0.039          & 0.842 ± 0.081          \\
GIN                     & 0.654 ± 0.032          & 0.679 ± 0.035          & 0.843 ± 0.082          \\
SIB                     & 0.577 ± 0.032          & 0.627 ± 0.044          & 0.811 ± 0.090          \\
DIR-GNN                 & 0.644 ± 0.018          & 0.670 ± 0.027          & 0.839 ± 0.065          \\
ProtGNN                 & 0.610 ± 0.021          & 0.653 ± 0.031          & 0.843 ± 0.050          \\
BrainGNN                & 0.602 ± 0.030          & 0.627 ± 0.021          & 0.822 ± 0.085          \\
IBGNN                   & 0.630 ± 0.027          & 0.640 ± 0.031          & 0.848 ± 0.086          \\
CI-GNN                  & 0.668 ± 0.040          & 0.675 ± 0.033          & 0.875 ± 0.068          \\
\textbf{BrainIB (Ours)} & \textbf{0.700 ± 0.022} & \textbf{0.702 ± 0.020} & \textbf{0.903 ± 0.046} \\ \bottomrule
\end{tabular}}
\label{tab:ten-fold}
\end{table}

\subsubsection{Leave-One-Site-Out Cross Validation}
To further assess the generalization capacity of BrainIB to unseen data, we further implement leave-one-site-out cross validation.  ABIDE and REST-meta-MDD datasets contain $17$ independent sites. In detail, we divide each dataset into the training set ($16$ sites out of $17$ sites) to train the model, and the testing set (remaining site out of $17$ sites) for testing a model.  Additionally, we compare the BrainIB with three SOTA models including GIN, DIR-GNN and CI-GNN for performance evaluation, since GIN, DIR-GNN and CI-GNN perform better in tenfold cross validation than other SOTA models.

For REST-meta-MDD, age, education, Hamilton Depression Scale (HAMD), illness duration, gender and sample size are selected as six representative factors. HAMD are clinical profiles for depression symptoms. The lower value of the HAMD indicates the less severity of depression. The experimental results are summarized in Table \ref{tab:Leave-one-set-out-MDD}. Compared with the baselines, BrainIB achieves the highest mean generalization accuracy of 71.8\% and yields significant and consistent improvements in most sites, suggesting an strength of our study that the BrainIB could generalize well on the completely different multi-sites even if site-specific difference. To be specific, the most classifiers achieved generalization accuracy of more than 70\%, even 85.4\% for site 3.  Compared with other SOTA deep models such as ProtGNN, BrainIB achieves with up to more than 17.1\% absolute improvements in site 3. Interestingly, we observe the values of age and education in site 3 are similar with site 1 and site 5, suggesting that the better generalization performance in site 3 result from the prediction of homogeneous data (site 1 and site 5). However, classifiers developed in site 1 and site 14  achieved generalization accuracy of less than 70\%. Similarly, we speculate that the poor generalization performance of site 1 and site 14 result from the prediction of heterogenous data. The values of representative factors in site 1 are different from the most sites, while sites with similar factors such as site 5 often obtain the poor generalization performance.

\begin{table*}[ht]
\caption{ Leave-one-set-out cross validation on Rest-meta-MDD. All the performance of  methods are reported under their best settings.}
\begin{threeparttable}
\renewcommand\arraystretch{1}
\resizebox{\linewidth}{!}{
\begin{tabular}{@{}ccccccc|cccccc|cc@{}}
\toprule[1.5pt]
\multirow{2}{*}{\textbf{Site}} & \multirow{2}{*}{\textbf{Age(y)}} & \multirow{2}{*}{\textbf{Education(y)}} & \multirow{2}{*}{\textbf{HAMD}} & \multirow{2}{*}{\textbf{Illness Duration(m)}} & \multirow{2}{*}{\textbf{Gender(M/F)}} & \multirow{2}{*}{\textbf{Sample(MDD/HC)}} & \multicolumn{2}{c}{\textbf{GIN}} & \multicolumn{2}{c}{\textbf{DIR-GNN}} & \multicolumn{2}{c|}{\textbf{CI-GNN}} & \multicolumn{2}{c}{\textbf{BrainIB}} \\ \cmidrule(l){8-15} 
                               &                                  &                                        &                                &                                                &                                       &                                          & \textbf{Acc}    & \textbf{Sen}   & \textbf{Acc}      & \textbf{Sen}     & \textbf{Acc}      & \textbf{Sen}     & \textbf{Acc}      & \textbf{Sen}     \\ 
  \midrule[0.8pt]
{\color[HTML]{000000} site1} &
  {\color[HTML]{000000} 31.8 ± 8.5} &
  {\color[HTML]{000000} 14.5 ± 2.7} &
  {\color[HTML]{000000} 24.9 ± 4.8} &
  {\color[HTML]{000000} 5.3 ± 5.1} &
  {\color[HTML]{000000} 62/84} &
  {\color[HTML]{000000} 73/73} &
  {\color[HTML]{000000} 58.2\%} &
  65.8\% &
  56.8\% &
  {\color[HTML]{000000} 75.3\%} &
  \textbf{63.3\%} &
  54.3\% &
  \textbf{63.3\%} &
  \textbf{79.5\%} \\
{\color[HTML]{000000} site2} &
  {\color[HTML]{000000} 43.5 ± 11.7} &
  {\color[HTML]{000000} 10.8 ± 4.6} &
  {\color[HTML]{000000} 22.9 ± 3.0} &
  {\color[HTML]{000000} 29.8 ± 28.5} &
  {\color[HTML]{000000} 5/25} &
  {\color[HTML]{000000} 16/14} &
  {\color[HTML]{000000} 66.7\%} &
  {\color[HTML]{000000} 71.4\%} &
  70.0\% &
  82.6\% &
  {\color[HTML]{000000} \textbf{83.0\%}} &
  73.9\% &
  73.0\% &
  \textbf{85.7\%}\\
{\color[HTML]{000000} site3} &
  {\color[HTML]{000000} 30.0 ± 6.2} &
  {\color[HTML]{000000} 14.6 ± 3.1} &
  {\color[HTML]{000000} \textit{-}} &
  {\color[HTML]{000000} \textit{-}} &
  {\color[HTML]{000000} 19/22} &
  {\color[HTML]{000000} 18/23} &
  {\color[HTML]{000000} 80.5\%} &
  {\color[HTML]{000000} \textbf{82.6\%}} &
  78.0\% &
  60.8\% &
  {\color[HTML]{000000} 85.2\%} &
  81.1\% &
  \textbf{85.4\%} &
  \textbf{82.6\%} \\
{\color[HTML]{000000} site4} &
  {\color[HTML]{000000} 40.0 ± 11.8} &
  {\color[HTML]{000000} 13.1 ± 4.5} &
  {\color[HTML]{000000} 22.1 ± 4.4} &
  {\color[HTML]{000000} 41.5 ± 43.2} &
  {\color[HTML]{000000} 27/45} &
  {\color[HTML]{000000} 35/37} &
  {\color[HTML]{000000} 72.2\%} &
  {\color[HTML]{000000} 78.3\%} &
  75.0\% &
  86.5\% &
  {\color[HTML]{000000} 76.3\%}&
    75.7\% &
  \textbf{\textbf{77.8\%}} &
  \textbf{\textbf{94.6\%}} \\
{\color[HTML]{000000} site5} &
  {\color[HTML]{000000} 31.9 ± 10.3} &
  {\color[HTML]{000000} 12.1 ± 3.2} &
  {\color[HTML]{000000} 24.3 ± 7.9} &
  {\color[HTML]{000000} 16.6 ± 22.9} &
  {\color[HTML]{000000} 30/57} &
  {\color[HTML]{000000} 39/48} &
  {\color[HTML]{000000} 71.3\%} &
  {\color[HTML]{000000} \textbf{68.8\%}} &
  64.4\% &
  72.9\% &
  {\color[HTML]{000000} \textbf{70.0\%}}&
    68.8\% &
  \textbf{71.3\%} &
  \textbf{75.0\%} \\
{\color[HTML]{000000} site6} &
  {\color[HTML]{000000} 28.6 ± 8.3} &
  {\color[HTML]{000000} 14.7 ± 3.1} &
  {\color[HTML]{000000} -} &
  {\color[HTML]{000000} 31.3 ± 43.1} &
  {\color[HTML]{000000} 52/44} &
  {\color[HTML]{000000} 48/48} &
  {\color[HTML]{000000} 66.7\%} &
  {\color[HTML]{000000} 56.3\%} &
  \textbf{68.8\%} &
  64.6\% &
  {\color[HTML]{000000} \textbf{68.8\%}} &
  68.8\% &
  \textbf{68.8\%} &
  \textbf{75.0\%} \\
{\color[HTML]{000000} site7} &
  {\color[HTML]{000000} 32.7 ± 9.8} &
  {\color[HTML]{000000} 11.9 ± 2.8} &
  {\color[HTML]{000000} 20.7 ± 3.7} &
  {\color[HTML]{000000} 11.6 ± 23.0} &
  {\color[HTML]{000000} 38/33} &
  {\color[HTML]{000000} 45/26} &
  {\color[HTML]{000000} 73.2\%} &
  {\color[HTML]{000000} 73.1\%} &
  70.4\% &
  84.6\% &
  {\color[HTML]{000000} \textbf{75.4\%}} &
  69.2\% &
  73.2\% &
  \textbf{88.5\%} \\
{\color[HTML]{000000} site8} &
  {\color[HTML]{000000} 30.8 ± 9.3} &
  {\color[HTML]{000000} 13.2 ± 3.6} &
  {\color[HTML]{000000} 21.6 ± 3.2} &
  {\color[HTML]{000000} 29.2 ± 23.9} &
  {\color[HTML]{000000} 17/20} &
  {\color[HTML]{000000} 20/17} &
  {\color[HTML]{000000} 67.6\%} &
  {\color[HTML]{000000} 64.7\%} &
  67.6\% &
  82.4\% &
  {\color[HTML]{000000} \textbf{73.2\%}} &
  \textbf{88.2\%} &
  \textbf{75.7\%} &
  81.3\% \\
{\color[HTML]{000000} site9} &
  {\color[HTML]{000000} 33.4 ± 9.5} &
  {\color[HTML]{000000} 13.5 ± 2.2} &
  {\color[HTML]{000000} 24.8 ± 3.9} &
  {\color[HTML]{000000} -} &
  {\color[HTML]{000000} 13/23} &
  {\color[HTML]{000000} 20/16} &
  {\color[HTML]{000000} 72.2\%} &
  {\color[HTML]{000000} 75.0\%} &
  \textbf{80.6\%} &
  81.3\% &
  {\color[HTML]{000000} 72.2\%} &
  81.3\% &
  \textbf{80.6\%} &
  \textbf{93.8\%} \\
{\color[HTML]{000000} site10} &
  {\color[HTML]{000000} 29.9 ± 6.3} &
  {\color[HTML]{000000} 14.0 ± 3.2} &
  {\color[HTML]{000000} 21.1 ± 3.2} &
  {\color[HTML]{000000} 6.1 ± 4.1} &
  {\color[HTML]{000000} 34/59} &
  {\color[HTML]{000000} 61/32} &
  {\color[HTML]{000000} 71.0\%} &
  {\color[HTML]{000000} 59.4\%} &
   71.0\% &
   71.9\% &
  {\color[HTML]{000000} 68.3\%} &
    53.1\% &
  \textbf{72.0\%} &
  \textbf{78.1\%}\\
{\color[HTML]{000000} site11} &
  {\color[HTML]{000000} 42.8 ± 14.1} &
  {\color[HTML]{000000} 12.2 ± 3.9} &
  {\color[HTML]{000000} 26.5 ± 5.3} &
  {\color[HTML]{000000} 41.0 ± 57.2} &
  {\color[HTML]{000000} 26/41} &
  {\color[HTML]{000000} 30/37} &
  {\color[HTML]{000000} 79.1\%} &
  {\color[HTML]{000000} 86.5\%} &
   70.0\% &
   68.3\% &
  {\color[HTML]{000000} 81.1\%} &
   89.2\% &
   \textbf{82.1\%} &
  \textbf{94.6\%}\\
{\color[HTML]{000000} site12} &
  {\color[HTML]{000000} 21.2 ± 14.1} &
  {\color[HTML]{000000} 12.2 ± 3.9} &
  {\color[HTML]{000000} 20.7 ± 5.6} &
  {\color[HTML]{000000} \textit{-}} &
  {\color[HTML]{000000} 27/55} &
  {\color[HTML]{000000} 41/41} &
  {\color[HTML]{000000} 59.8\%} &
  {\color[HTML]{000000} 70.7\%} &
   64.6\% &
   75.0\% &
  {\color[HTML]{000000} 62.3\%} &
   51.6\% &
   \textbf{67.1\%} &
  \textbf{80.5\%}\\
{\color[HTML]{000000} site13} &
  {\color[HTML]{000000} 35.1 ± 10.6} &
  {\color[HTML]{000000} 9.8 ± 3.6} &
  {\color[HTML]{000000} 19.4 ± 8.9} &
  {\color[HTML]{000000} 87.0 ± 99.6} &
  {\color[HTML]{000000} 19/30} &
  {\color[HTML]{000000} 18/31} &
  {\color[HTML]{000000} 67.3\%} &
  {\color[HTML]{000000} 58.1\%} &
   67.3\% &
   \textbf{93.5\%} &
  {\color[HTML]{000000} 67.3\%} &
   71.0\% &
   \textbf{69.4\%} &
  76.9\%\\
{\color[HTML]{000000} site14} &
  {\color[HTML]{000000} 38.9 ± 13.8} &
  {\color[HTML]{000000} 12.0 ± 3.7} &
  {\color[HTML]{000000} 21.0 ± 5.7} &
  {\color[HTML]{000000} 50.8 ± 64.6} &
  {\color[HTML]{000000} 149/321} &
  {\color[HTML]{000000} 245/225} &
  {\color[HTML]{000000} 60.6\%} &
  {\color[HTML]{000000} 47.1\%} &
   56.6\% &
   63.1\% &
  {\color[HTML]{000000} \textbf{63.2\%}} &
   64.6\% &
   \textbf{63.2\%} &
  \textbf{75.4\%}\\
{\color[HTML]{000000} site15} &
  {\color[HTML]{000000} 35.2 ± 12.3} &
  {\color[HTML]{000000} 12.3 ± 2.5} &
  {\color[HTML]{000000} 14.3 ± 8.1} &
  {\color[HTML]{000000} 86.7 ± 92.7} &
  {\color[HTML]{000000} 62/82} &
  {\color[HTML]{000000} 79/65} &
  {\color[HTML]{000000} 64.6\%} &
  {\color[HTML]{000000} 46.2\%} &
   61.1\% &
   50.8\% &
  {\color[HTML]{000000} 68.3\%} &
   61.5\% &
   \textbf{70.1\%}&
   \textbf{80.0\%}\\
{\color[HTML]{000000} site16} &
  {\color[HTML]{000000} 28.8 ± 9.6} &
  {\color[HTML]{000000} 12.7 ± 2.6} &
  {\color[HTML]{000000} 23.1 ± 4.3} &
  {\color[HTML]{000000} 28.7 ± 26.6} &
  {\color[HTML]{000000} 21/17} &
  {\color[HTML]{000000} 18/20} &
  {\color[HTML]{000000} 65.8\%} &
  {\color[HTML]{000000} 65.0\%} &
   68.4\% &
   65.0\% &
  {\color[HTML]{000000} \textbf{75.2\%}} &
   \textbf{75.0\%} &
   71.1\%&
   70.0\%\\
{\color[HTML]{000000} site17} &
  {\color[HTML]{000000} 29.7 ± 10.5} &
  {\color[HTML]{000000} 14.1 ± 3.6} &
  {\color[HTML]{000000} 18.5 ± 8.7} &
  {\color[HTML]{000000} 19.5 ± 26.1} &
  {\color[HTML]{000000} 18/27} &
  {\color[HTML]{000000} 22/23} &
  {\color[HTML]{000000} 64.4\%} &
  {\color[HTML]{000000} \textbf{78.3\%}} &
   \textbf{68.9\%} &
   60.9\% &
  {\color[HTML]{000000} 64.4\%} &
   52.2\% &
   \textbf{68.9\%}&
   \textbf{78.3\%}\\
{\color[HTML]{000000} Mean} &
  {\color[HTML]{000000} 34.4 ± 12.3} &
  {\color[HTML]{000000} 12.8 ± 3.5} &
  {\color[HTML]{000000} 21.2 ± 6.5} &
  {\color[HTML]{000000} 39.1 ± 61.1} &
  {\color[HTML]{000000} 36/58} &
  {\color[HTML]{000000} 49/46} &
  {\color[HTML]{000000} 68.3\%} &
  {\color[HTML]{000000} 67.5\%} &
   68.2\% &
  {\color[HTML]{000000} 72.9\%} &
   71.6\% &
   70.0\%&
   \textbf{72.5\%}&
   \textbf{81.8\%}\\
  \bottomrule[1.5pt]
\end{tabular}}
\begin{tablenotes}
 \footnotesize
 \item  {``-'' denotes the missing values; ACC, Accuracy; Sen, Sensitivity.}
\end{tablenotes}
\end{threeparttable}
\label{tab:Leave-one-set-out-MDD}
\end{table*}

For ABIDE, MRI vendors, age, FIQ (Full-Scale Intelligence Quotient), gender and sample size are selected as five representative factors. FIQ is able to reflect the degree of brain development. Table \ref{tab:Leave-one-set-out-ASD} shows the generalized performance of BrainIB and phenotypic information for the 17 sites. As can be seen, BrainIB outperforms four SOTA deep models in mean generalization accuracy by large margins,  with up to more than 2\% absolute improvements. Specifically, the most classifiers achieved generalization accuracy of more than 70\%, even 83.3\% for CMU. However, four sites demonstrates significantly reduced accuracies than the accuracy of 70\% : MAX\_MUN, OHSU, PITT and TRINITY, which are consistent with the previous studies \cite{heinsfeld2018identification, huang2020identifying}. These results suggest that these centers might be provided with site-specific variability and heterogeneity, which leads to the poor generalization performance. Additionally, we also utilize sensitivity as a measure of the model's performance in addition to accuracy. Sensitivity, also known as true positive rate, measures the percentage of correctly identified cases among those who actually are psychiatry disorder, and reflects the model's ability to detect patients. BrainIB achieves better sensitivity for the most sites on both datasets, indicating that brainIB has a better ability to identify patients. Therefore, BrainIB is more robust to generalization to a new different site and outperforms the baselines on classification performance by a large margin. BrainIB is able to provide clear diagnosis boundary which allows accurate diagnosis of MDD and ASD and ignores the effects of site. 

\begin{table*}[ht]
\center
\caption{ Leave-one-set-out cross validation on ABIDE. All the performance of  methods are reported under their best settings.}
\center
\renewcommand\arraystretch{1}
\resizebox{\linewidth}{!}{\begin{tabular}{clcccc|cccccc|cc}
\hline
\toprule[1.5pt]
\multirow{2}{*}{\textbf{Site}} & \multirow{2}{*}{\textbf{Vendor}} & \multirow{2}{*}{\textbf{Age(y)}} & \multirow{2}{*}{\textbf{FIQ}} & \multirow{2}{*}{\textbf{Gender(M/F)}} & \multirow{2}{*}{\textbf{Sample(MDD/HC)}} & \multicolumn{2}{c}{\textbf{GIN}} & \multicolumn{2}{c}{\textbf{DIR-GNN}} & \multicolumn{2}{c|}{\textbf{CI-GNN}} & \multicolumn{2}{c}{\textbf{BrainIB}} \\ \cline{7-14} 
                               &                                  &                                  &                               &                                       &                                          & \textbf{Acc}    & \textbf{Sen}   & \textbf{Acc}      & \textbf{Sen}     & \textbf{Acc}      & \textbf{Sen}     & \textbf{Acc}      & \textbf{Sen}     \\ 
  \midrule[0.8pt]
{\color[HTML]{000000} CMU} &
  {\color[HTML]{000000} Siemens} &
  {\color[HTML]{000000} 26.6 ± 5.8} &
  {\color[HTML]{000000} 115.3 ± 9.8} &
  {\color[HTML]{000000} 18/6} &
  {\color[HTML]{000000} 11/13} &
  {\color[HTML]{000000} \textbf{83.3\%}} &
  {\color[HTML]{000000} 90.0\%} &
   \textbf{83.3\%} &
   84.6\% &
   \textbf{83.3\%} &
   90.0\% &
   \textbf{83.3\%} &
  {\color[HTML]{000000} \textbf{100.0\%}} \\
{\color[HTML]{000000} CALTECH} &
  {\color[HTML]{000000} Siemens} &
  {\color[HTML]{000000} 28.2 ± 10.7} &
  {\color[HTML]{000000} 111.3 ± 11.4} &
  {\color[HTML]{000000} 30/8} &
  {\color[HTML]{000000} 19/19} &
  {\color[HTML]{000000} \textbf{71.1\%}} &
  {\color[HTML]{000000} 66.7\%} &
   68.4\% &
   57.6\% &
  \textbf{71.1\%} &
   73.3\% &
    \textbf{71.1\%} &
  {\color[HTML]{000000} \textbf{73.7\%}} \\
{\color[HTML]{000000} KKI} &
  {\color[HTML]{000000} Philips} &
  {\color[HTML]{000000} 10.1 ± 1.3} &
  {\color[HTML]{000000} 106.7 ± 15.3} &
  {\color[HTML]{000000} 42/13} &
  {\color[HTML]{000000} 22/33} &
  {\color[HTML]{000000} 70.9\%} &
  {\color[HTML]{000000} 65.5\%} &
   65.5\% &
   68.6\% &
   67.4\% &
   65.5\% &
   \textbf{72.7\%} &
  {\color[HTML]{000000} \textbf{71.4\%}} \\
{\color[HTML]{000000} LEUVEN} &
  {\color[HTML]{000000} Philips} &
  {\color[HTML]{000000} 18.0 ± 5.0} &
  {\color[HTML]{000000} 112.2 ± 13.0} &
  {\color[HTML]{000000} 56/8} &
  {\color[HTML]{000000} 29/35} &
  {\color[HTML]{000000} \textbf{73.4\%}} &
  {\color[HTML]{000000} 72.5\%} & 
   68.8\% &
   71.4\% &
   67.8\% &
   72.5\% &
   \textbf{73.4\%} &
  {\color[HTML]{000000} \textbf{94.3\%}} \\
{\color[HTML]{000000} MAX\_MUN} &
  {\color[HTML]{000000} Siemens} &
  {\color[HTML]{000000} 26.2 ± 12.1} &
  {\color[HTML]{000000} 110.8 ± 11.3} &
  {\color[HTML]{000000} 50/7} &
  {\color[HTML]{000000} 24/33} &
  {\color[HTML]{000000} 64.9\%} &
  {\color[HTML]{000000} 87.9\%} &
   64.9\% &
   \textbf{87.9\%} &
   \textbf{71.7\%} &
   76.0\% &
   66.7\% &
  {\color[HTML]{000000} \textbf{87.9\%}} \\
{\color[HTML]{000000} NYU} &
  {\color[HTML]{000000} Siemens} &
  {\color[HTML]{000000} 15.3 ± 6.6} &
  {\color[HTML]{000000} 110.9 ± 14.9} &
  {\color[HTML]{000000} 147/37} &
  {\color[HTML]{000000} 79/105} &
  {\color[HTML]{000000} 69.0\%} &
  {\color[HTML]{000000} 70.2\%} &
   63.6\% &
   81.0\% &
   64.3\% &
   64.3\% &
   \textbf{70.1\%} &
  {\color[HTML]{000000} \textbf{89.5\%}} \\
{\color[HTML]{000000} OHSU} &
  {\color[HTML]{000000} Siemens} &
  {\color[HTML]{000000} 10.8 ± 1.9} &
  {\color[HTML]{000000} 111.6 ± 16.9} &
  {\color[HTML]{000000} 28/0} &
  {\color[HTML]{000000} 13/15} &
  {\color[HTML]{000000} \textbf{71.4\%}} &
  {\color[HTML]{000000} 71.4\%} &
  67.9\% &
  \textbf{73.3\%} &
  67.9\% &
  \textbf{73.3\%} &
  67.9\% &
  {\color[HTML]{000000} \textbf{73.3\%}} \\
{\color[HTML]{000000} OLIN} &
  {\color[HTML]{000000} Siemens} &
  {\color[HTML]{000000} 16.8 ± 3.5} &
  {\color[HTML]{000000} 113.9 ± 17.0} &
  {\color[HTML]{000000} 31/5} &
  {\color[HTML]{000000} 20/16} &
  {\color[HTML]{000000} 72.2\%} &
  {\color[HTML]{000000} 60.9\%} &
   \textbf{77.8\%}&
   75.0\% &
   \textbf{77.8\%}&
   \textbf{77.8\%} &
   \textbf{77.8\%} &
  {\color[HTML]{000000} 75.0\%} \\
{\color[HTML]{000000} PITT} &
  {\color[HTML]{000000} Siemens} &
  {\color[HTML]{000000} 18.9 ± 6.9} &
  {\color[HTML]{000000} 110.1 ± 12.2} &
  {\color[HTML]{000000} 49/18} &
  {\color[HTML]{000000} 30/27} &
  {\color[HTML]{000000} 66.7\%} &
  {\color[HTML]{000000} 73.3\%} &
   70.2\% &
  51.9\% &
  \textbf{78.1\%} &
  75.0\% &
  66.7\% &
  {\color[HTML]{000000} \textbf{88.9}\%} \\
{\color[HTML]{000000} SBL} &
  {\color[HTML]{000000} Philips} &
  {\color[HTML]{000000} 34.4 ± 8.6} &
  {\color[HTML]{000000} 109.2 ± 13.6} &
  {\color[HTML]{000000} 30/0} &
  {\color[HTML]{000000} 15/15} &
  {\color[HTML]{000000} 76.7\%} &
  {\color[HTML]{000000} 83.3\%} &
   76.7\% &
   66.7\% &
   69.2\% &
   83.3\% &
   \textbf{83.3\%} &
  {\color[HTML]{000000} \textbf{93.3\%}} \\
{\color[HTML]{000000} SDSU} &
  {\color[HTML]{000000} GE} &
  {\color[HTML]{000000} 14.4 ± 1.8} &
  {\color[HTML]{000000} 109.4 ± 13.8} &
  {\color[HTML]{000000} 29/7} &
  {\color[HTML]{000000} 14/22} &
  {\color[HTML]{000000} 75.0\%} &
  {\color[HTML]{000000} 83.3\%} &
   75.0\% &
   \textbf{90.9\%} &
   \textbf{83.3\%} &
   77.8\% &
   75.0\% &
  {\color[HTML]{000000} \textbf{90.9\%}} \\
{\color[HTML]{000000} STANFORD} &
  {\color[HTML]{000000} GE} &
  {\color[HTML]{000000} 10.0 ± 1.6} &
  {\color[HTML]{000000} 112.3 ± 16.4} &
  {\color[HTML]{000000} 32/8} &
  {\color[HTML]{000000} 20/20} &
  {\color[HTML]{000000} 67.5\%} &
  {\color[HTML]{000000} 77.3\%} &
   \textbf{75.0\%} &
   70.0\% &
   \textbf{75.0\%} &
   72.7\% &
   \textbf{75.0\%} &
  {\color[HTML]{000000} \textbf{84.0\%}} \\
{\color[HTML]{000000} TRINITY} &
  {\color[HTML]{000000} Philips} &
  {\color[HTML]{000000} 17.2 ± 3.6} &
  {\color[HTML]{000000} 110.1 ± 13.5} &
  {\color[HTML]{000000} 49/0} &
  {\color[HTML]{000000} 24/25} &
  {\color[HTML]{000000} 63.3\%} &
  {\color[HTML]{000000} 66.7\%} &
   \textbf{67.3\%} &
   \textbf{72.0\%} &
    64.2\% &
   70.0\% &
   65.3\% &
  {\color[HTML]{000000} \textbf{72.0\%}} \\
{\color[HTML]{000000} UCLA} &
  {\color[HTML]{000000} Siemens} &
  {\color[HTML]{000000} 13.0 ± 2.2} &
  {\color[HTML]{000000} 103.2 ± 12.8} &
  {\color[HTML]{000000} 87/12} &
  {\color[HTML]{000000} 54/45} &
  {\color[HTML]{000000} 66.7\%} &
  {\color[HTML]{000000} 71.4\%} &
   68.7\% &
   68.9\% &
   63.4\% &
   66.7\% &
   \textbf{74.7\%} &
  {\color[HTML]{000000} \textbf{80.0\%}} \\
{\color[HTML]{000000} UM} &
  {\color[HTML]{000000} GE} &
  {\color[HTML]{000000} 14.0 ± 3.2} &
  {\color[HTML]{000000} 106.9 ± 14.0} &
  {\color[HTML]{000000} 117/28} &
  {\color[HTML]{000000} 68/77} &
  {\color[HTML]{000000} 64.1\%} &
  {\color[HTML]{000000} 58.4\%} &
   \textbf{66.2\%} &
   77.9\% &
   61.2\% &
   75.3\% &
   64.8\% &
  {\color[HTML]{000000} \textbf{79.2\%}} \\
{\color[HTML]{000000} USM} &
  {\color[HTML]{000000} Siemens} &
  {\color[HTML]{000000} 22.1 ± 7.7} &
  {\color[HTML]{000000} 106.6 ± 16.7} &
  {\color[HTML]{000000} 101/0} &
  {\color[HTML]{000000} 58/43} &
  {\color[HTML]{000000} 71.3\%} &
  {\color[HTML]{000000} 75.4\%} &
   67.3\% &
  58.1\% &
  66.3\% &
  67.7\% &
  \textbf{73.3\%} &
  {\color[HTML]{000000} \textbf{86.0\%}} \\
{\color[HTML]{000000} YALE} &
  {\color[HTML]{000000} Siemens} &
  {\color[HTML]{000000} 12.7 ± 2.9} &
  {\color[HTML]{000000} 99.8 ± 20.1} &
  {\color[HTML]{000000} 40/16} &
  {\color[HTML]{000000} 28/28} &
  {\color[HTML]{000000} 78.6\%} &
  {\color[HTML]{000000} 76.2\%} &
   73.2\% &
   78.6\% &
   75.1\% &
   80.9\% &
   \textbf{82.1\%} &
  {\color[HTML]{000000} \textbf{82.1\%}} \\
{\color[HTML]{000000} Mean} &
  {\color[HTML]{000000} N.A.} &
  {\color[HTML]{000000} 17.1 ± 8.1} &
  {\color[HTML]{000000} 108.5 ± 15.0} &
  {\color[HTML]{000000} 55/10} &
  {\color[HTML]{000000} 31/34} &
  {\color[HTML]{000000} 70.9\%} &
  {\color[HTML]{000000} 73.6\%} &
   70.6\% &
   72.6\% &
   71.0\% &
   74.2\% &
   \textbf{73.1\%} &
  {\color[HTML]{000000} \textbf{83.6\%}} \\ 
  \bottomrule[1.5pt]
\end{tabular}}
\begin{tablenotes}
 \footnotesize
 \item  {  ACC, Accuracy; Sen, Sensitivity.}
\end{tablenotes}
\label{tab:Leave-one-set-out-ASD}
\end{table*}

\subsection{Interpretation Analysis}
\subsubsection{Disease-Specific Brain Network Connections}
\begin{figure*}[ht!]
\centering
\includegraphics[scale=0.68]{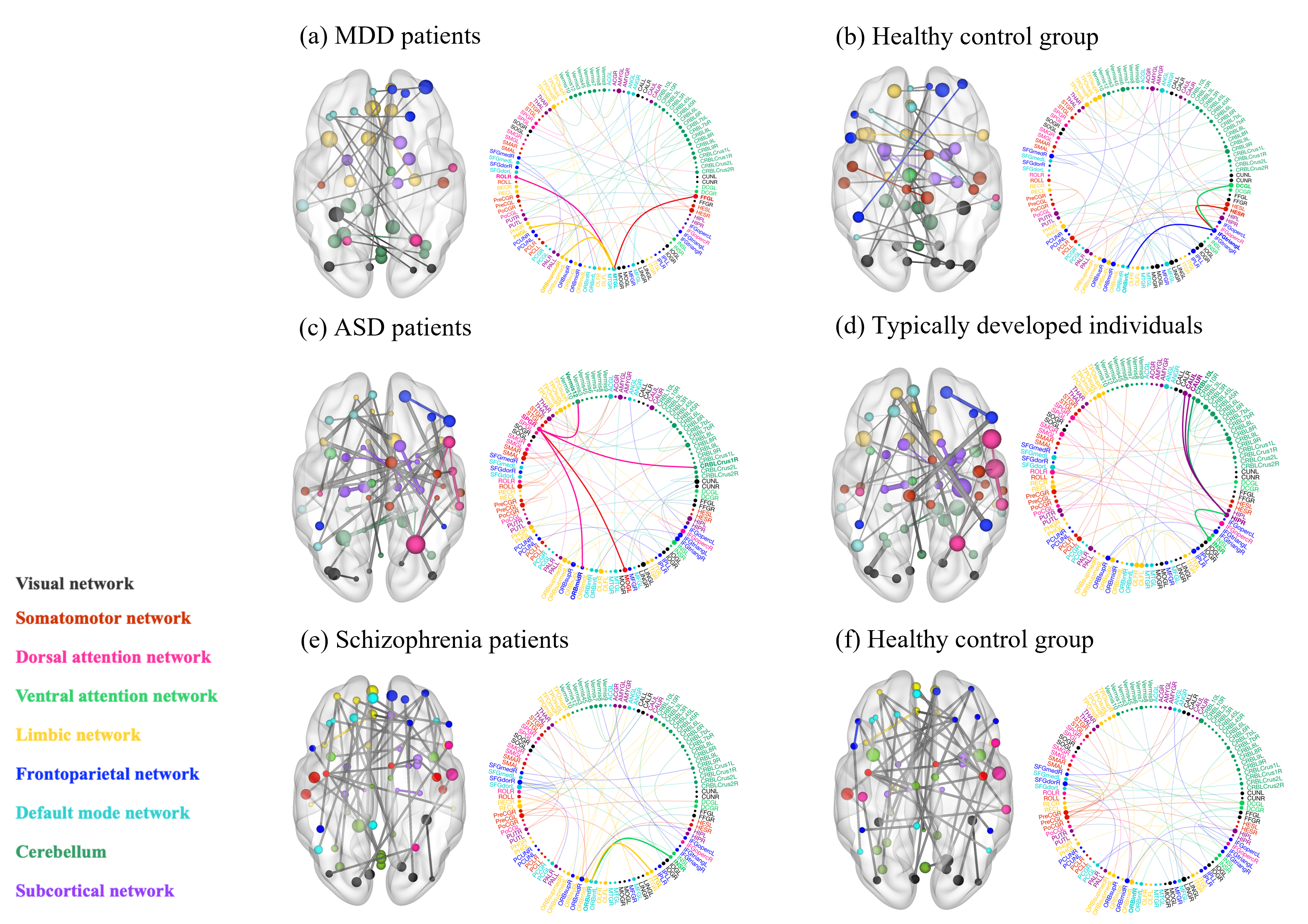}
\caption{Comparison of explanation graph connections in brain networks of healthy controls and patients on ASD, MDD and schizophrenia datasets. The colors of brain neural systems are described as: visual network, somatomotor network, dorsal attention network, ventral attention network, limbic network, frontoparietal network, default mode network, cerebellum and subcortial network respectively.}
\label{fig:connection}
\end{figure*}

In order to evaluate the interpretation of BrainIB, we further investigate the capability of IB-subgraph on interpreting the property of neural mechanism in patients with MDD and ASD. BrainIB could capture the important subgraph structure in each subject. To compare inter-group differences in subgraphs, we calculate the average subgraph and select the top 50 edges to generate the dominant subgraph. Figure~\ref{fig:connection} demonstrates the comparison of dominant subgraph $G_{dsub}$ for healthy  controls and patient groups on three datasets, in which the color of nodes represents distinct brain networks and the size of an edge is determined by its weight in the dominant subgraph. 

As can be seen, patients with MDD exhibits tight interactions between default mode network and limbic network, while these connections in healthy controls are much more sparse. These patterns are consistent with the findings in Korgaonkar et al. \cite{korgaonkar2020intrinsic}, where connections between default mode network and limbic network are a predictive biomarker of remission in major depressive disorder. Abnormal affective processing and maladaptive rumination are core feature of MDD \cite{belmaker2008major}. Our results found that DMN related to rumination \cite{kaiser2016dynamic} and limbic network associated with emotion processing \cite{rolls2015limbic}, providing an explanation for maladaptive rumination and negative emotion in MDD. In addition, connections within the frontoparietal system of patients are significantly less than that of healthy controls, which is consistent with the previous studies \cite{rai2021default}. For the ASD, patterns within dorsal attention network of patients are significantly more than that of typical development controls (TD), which is in line with a recent review \cite{mundy2018review}.  They review the hypothesis that the neurodevelopment of joint attention contributes to the development of neural systems for human social cognition. In addition, we also observe that the reduced number of connections within subcortical system in patients, which in line with a recent study~\cite{hoogman2022consortium}. In patients with schizophrenia, there is a higher occurrence of tight interactions within the default mode network (DMN), whereas in healthy controls, these connections are sparser. These findings are in line with a previous study, where increased functional connectivity within DMN is observed in patients with schizophrenia~\cite{woodward2011functional}. Furthermore, DMN is considered as a treatment biomarker for schizophrenia~\cite{hu2017review}. 

\subsubsection{Important Brain Systems }
To investigate the contributions of brain systems to the prediction of a specific disease, we further use graph theory analysis on the subgraphs to  discover important brain systems. Here, subgraph for each participant represent a weighted adjacency matrix, which doesn't remove low-probability edges. To exclude weak or irrelevant edges from graph analysis, we use network sparsity strategy to generate a series of connected networks with connection density ranging from 10\% to 50\% in increments of 10\%. For each connection density, six topological property measurements are estimated: (1) betweenness centrality (Bc) characterizes the effect of a node on information flow between other nodes. (2) degree centrality (Dc) reflects the information communication ability in the subgraph. (3) clustering coefficient (Cp) measures the likelihood the neighborhoods of a node are connected to each other. (4) nodal efficiency (Eff) characterizes the efficiency of parallel information transfer of a node in the subgraph. (5) local efficiency (LocEff) calculates how efficient the communication is among the first neighbors of a node when it is deleted. (6) shortest path length (Lp) quantifies the mean distance or routing efficiency between individual node and all the other nodes in the subgraph. More details are described as the previous studies \cite{mazrooyisebdani2020graph}. Then we compute the AUC (area under curve) of all topological property measurements of each node in the subgraph for each subject. Finally, all topological property measurements of 9 brain systems by averaging the  measurements of nodes assigned to the same brain systems. The experimental results are summarized in Table~\ref{fig:system}.

As shown in Table~\ref{fig:system}, we observed that the importance of somatomotor network (SMN) for healthy controls (HC) is stronger than that of patients with MDD, while default mode network (DMN) is more important for patients than HC. These observations is in line with the findings in Yang et al. \cite{yang2021disrupted}, where patients with MDD are characterized by decreased degree centrality in the somatomotor network. This is in line with our findings on Rest-meta-MDD in Fig.~\ref{fig:connection}, in which the excessive connected structure within DMN in patients with MDD. Moreover, we also observe the importance of attention network in ASD, which is similar with  the observations of explanition graph connections. 

\begin{table}[ht]
\center
\caption{Top ranked neural systems of the explanation subgraph on MDD and ASD for both Healthy Control (HC), typically developed (TD) individuals and Patient under six comparative measures.}
\renewcommand\arraystretch{1}
\begin{tabular}{@{}
>{\columncolor[HTML]{FFFFFF}}c 
>{\columncolor[HTML]{FFFFFF}}c 
>{\columncolor[HTML]{FFFFFF}}c 
>{\columncolor[HTML]{FFFFFF}}c 
>{\columncolor[HTML]{FFFFFF}}c @{}}
\toprule[1.5pt]
\cellcolor[HTML]{FFFFFF}{\color[HTML]{000000} } &
  \multicolumn{2}{c}{\cellcolor[HTML]{FFFFFF}{\color[HTML]{000000} \textbf{MDD Datasets}}} &
  \multicolumn{2}{c}{\cellcolor[HTML]{FFFFFF}{\color[HTML]{000000} \textbf{ASD Dataset}}} \\ \cmidrule(l){2-5} 
\multirow{-2}{*}{\cellcolor[HTML]{FFFFFF}{\color[HTML]{000000} \textbf{Measures}}} &
  {\color[HTML]{000000} MDD} &
  {\color[HTML]{000000} HC} &
  {\color[HTML]{000000} ASD} &
  {\color[HTML]{000000} TD} \\ \midrule[0.8pt]
{\color[HTML]{000000} Bc}     & {\color[HTML]{000000} SMN DMN} & {\color[HTML]{000000} SMN DMN} & {\color[HTML]{000000} VAN CBL} & {\color[HTML]{000000} VAN SMN} \\
{\color[HTML]{000000} Dc}     & {\color[HTML]{000000} DMN SMN} & {\color[HTML]{000000} SMN DMN} & {\color[HTML]{000000} CBL VN}  & {\color[HTML]{000000} LN DAN}  \\
{\color[HTML]{000000} Cp}     & {\color[HTML]{000000} SMN DMN} & {\color[HTML]{000000} SMN DMN} & {\color[HTML]{000000} FPN LN}  & {\color[HTML]{000000} VAN DMN} \\
{\color[HTML]{000000} Eff}    & {\color[HTML]{000000} DMN SMN} & {\color[HTML]{000000} SMN DMN} & {\color[HTML]{000000} CBL VN}  & {\color[HTML]{000000} LN DAN}  \\
{\color[HTML]{000000} LocEff} & {\color[HTML]{000000} DMN SMN} & {\color[HTML]{000000} SMN DMN} & {\color[HTML]{000000} DMN LN}  & {\color[HTML]{000000} DMN SMN} \\
{\color[HTML]{000000} Lp}     & {\color[HTML]{000000} DMN SMN} & {\color[HTML]{000000} SMN DMN} & {\color[HTML]{000000} LN BLN}  & {\color[HTML]{000000} DMN VN}  \\ \bottomrule[1.5pt]
\end{tabular}
\label{fig:system}
\end{table}

\subsubsection{Quantitative Study}

\begin{figure}[ht!]
  \centering
  \subfloat[\footnotesize{Input data}]{\includegraphics[scale=0.26]{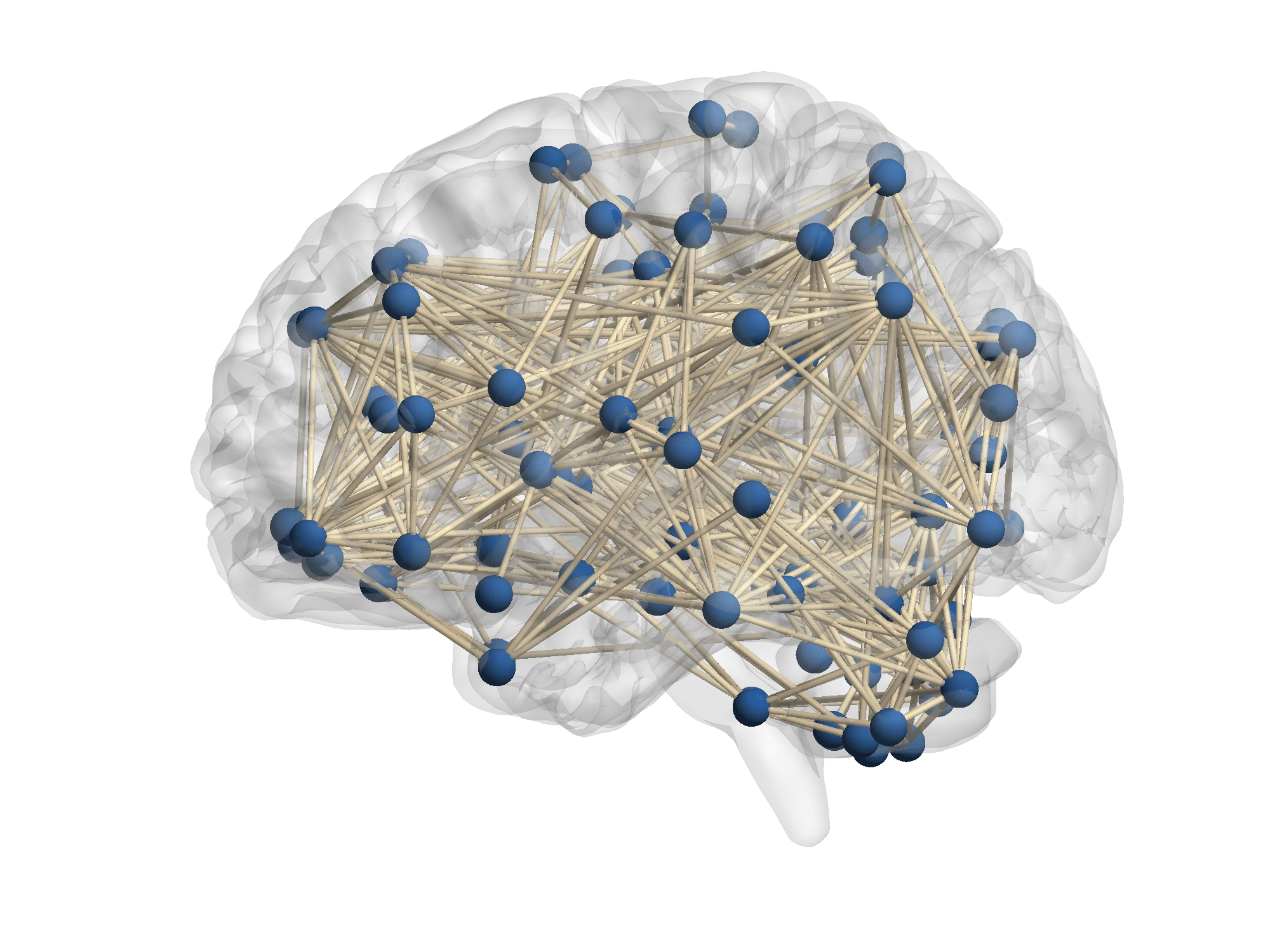}\label{fig:input}}
  \hfil
  \subfloat[\footnotesize{Perturbed data}]{\includegraphics[scale=0.26]{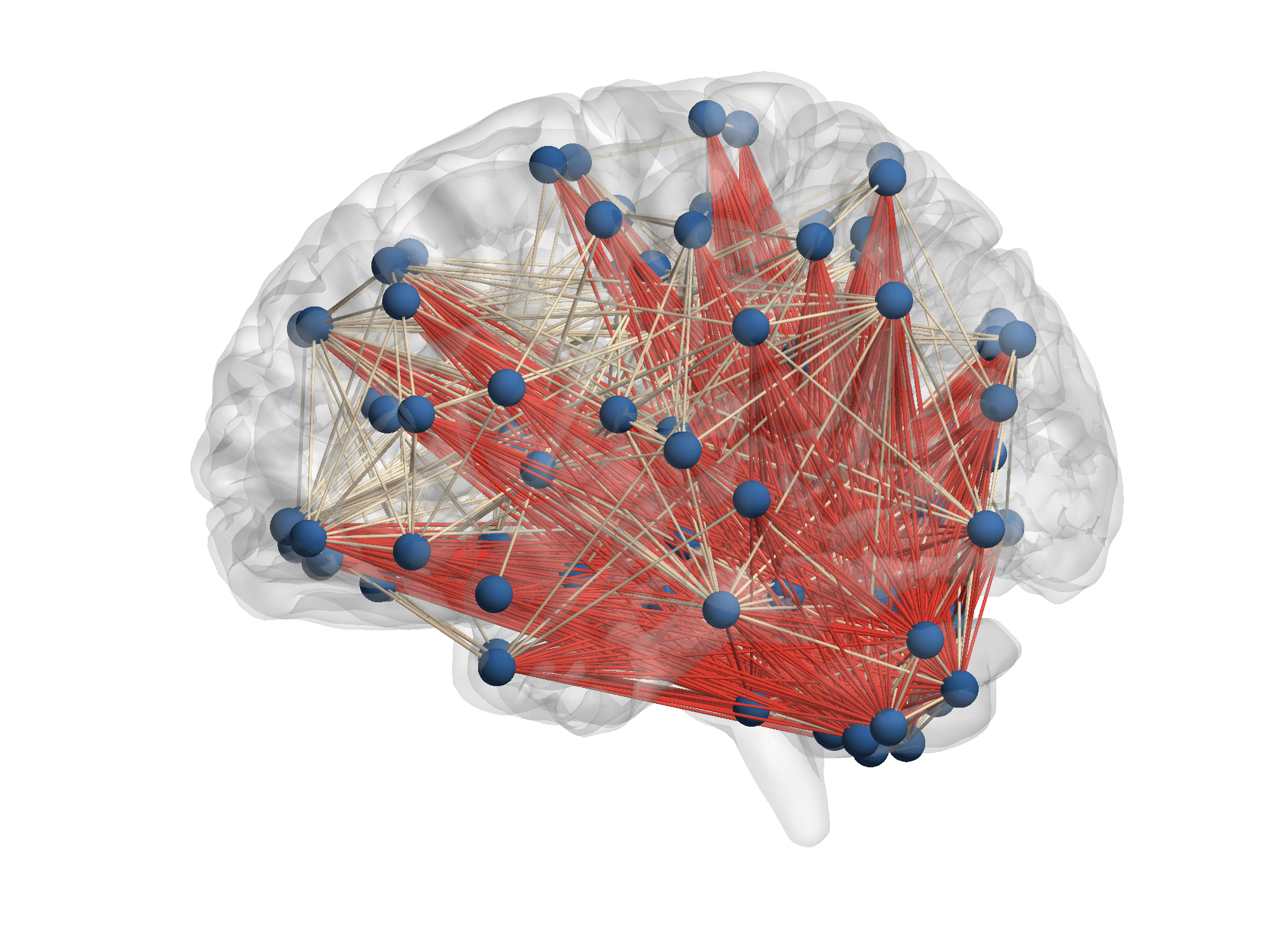}\label{fig:perturbeddata}}
  \hfil
  \subfloat[\footnotesize{ASD Patients}]{\includegraphics[scale=0.23]{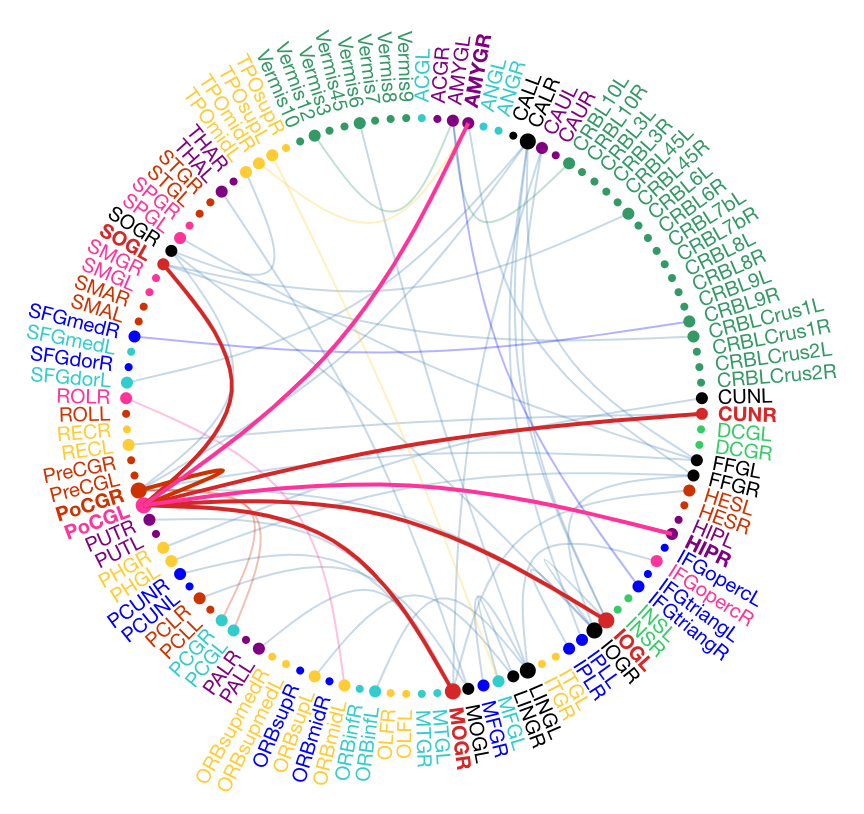}\label{fig:ASD_per}}
  \hfil
  \subfloat[\footnotesize{TD individuals}]{\includegraphics[scale=0.23]{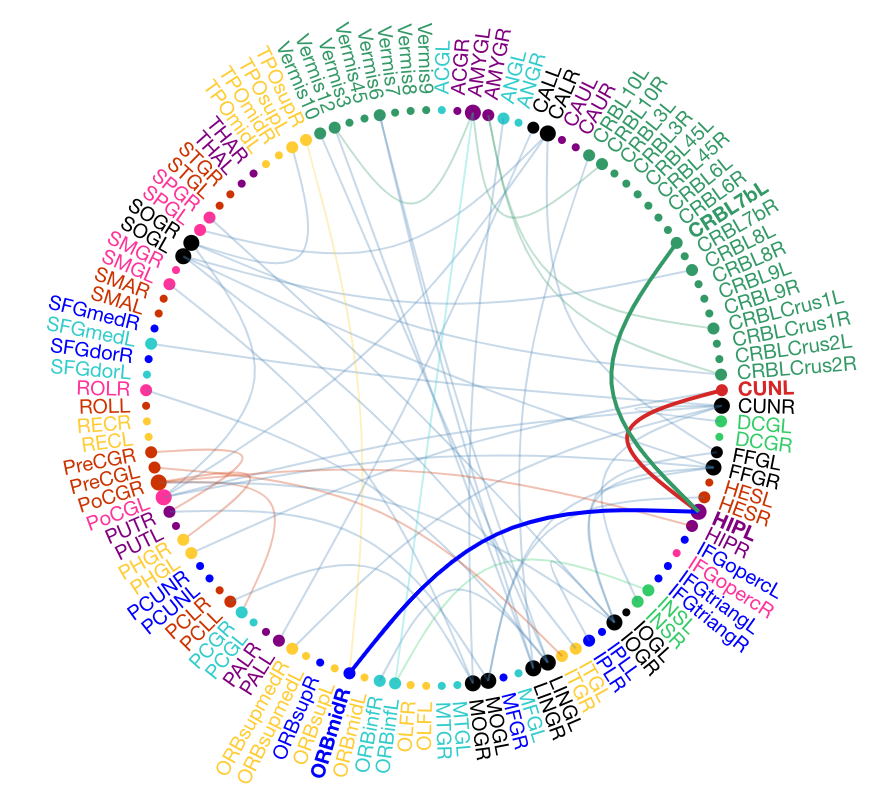}\label{fig:HC_per}}
  \hfil
  \caption{The influence of perturbed data on IB-subgraph using the ABIDE datasets. (a) Original input data. (b) Perturbed data where red edges are perturbed edges. Here, perturbed edges are cerebellum connections. (c) Subgraph connections in the patients with ASD using the Perturbed data. (d) Subgraph connections in the TD individuals using the Perturbed data.}
  \label{fig:perturbed}
\end{figure}

To further check the robust and interpretability of BrainIB, we compare the difference in predictions and subgraph structure obtained by BrainIB when feeding the original data and perturbed data. Specifically, we divide the ABIDE dataset into the training set (9 folds out of 10 folds), which is used for training a model, and the testing set (1 fold out of 10 folds) for testing a model. Then we construct perturbed data from testing set by adding irrelevant edges. Considering that ASD is a common psychiatric disorder characterised by early-onset difficulties in social communication~\cite{keller2022real} but cerebellum is a brain structure directly related to motor function~\cite{miterko2019consensus}, we add connections between and within cerebellum for the original data in testing set as perturbed data (see Fig.~\ref{fig:perturbed} (a)-(b)). Subsequently, we use training set to train the model and use respectively the original input data and the perturbed data to test the performance and interpretability of model. Results demonstrate that BrainIB with original data achieves the classification accuracy of 71.8\%, while BrainIB with perturbed data achieves the classification accuracy of 65.4\%, suggesting perturbed data could impair model performance. In addition, Fig.~\ref{fig:perturbed} (c)-(d) demonstrates subgraph from perturbed data for patients with ASD and typically developed (TD) individuals on ABIDE dataset. As can be seen, BrainIB successfully identify tight connections within dorsal attention network and reduced number of connections within subcortical network in patients with ASD, which is consistent with the results obtained with the original data. These results indicate that BrainIB is able to recognize disease-specific brain network connections with or without perturbing. 

\graphicspath{{./figs/}}

Moreover, we further assess if the selected edges obtained from BrainIB are truly relevant biomarkers. Specifically, we investigate the difference in classification accuracy between removing the edges of the identified subgraphs (BrainIB\_rest), randomly removing edges (BrainIB\_ran) and retaining only the edges of the identified subgraphs (BrainIB), in three different psychiatric datasets. According results are shown in Table~\ref{tab:edges}. Our findings show that removing edges of the identified subgraphs leads to a more significant decrease in BrainIB's performance compared to randomly removing edges no matter size of subgraph, providing evidence that these edges are indeed relevant biomarkers from the identified subgraphs.

\begin{table}[ht!]
\center
\caption{Changes in classification accuracy when removing the edges from the identified subgraphs, randomly removing edges, and not removing any edges. $K$ is the hyper-parameter which is able to determine the size of IB-subgraph.}
\renewcommand\arraystretch{1}
\begin{tabular}{@{}cccc@{}}
\toprule
\multicolumn{1}{l}{}  & \textbf{ABIDE}         & \textbf{REST-meta-MDD} & \textbf{SRPBS}         \\ \midrule
BrainIB\_rest ($K$ = 2) & 0.620 ± 0.037          & 0.627 ± 0.025          & 0.807 ± 0.071          \\
BrainIB\_ran ($K$ = 2)  & 0.640 ± 0.058          & 0.659 ± 0.039          & 0.842 ± 0.092          \\
BrainIB ($K$ = 2)       & \textbf{0.702 ± 0.020} & \textbf{0.700 ± 0.022} & \textbf{0.903 ± 0.046} \\ \hdashline
BrainIB\_rest ($K$ = 4) & 0.626 ± 0.026          & 0.625 ± 0.030          & 0.807 ± 0.089          \\
BrainIB\_ran ($K$ = 4)  & 0.652 ± 0.048          & 0.641 ± 0.048          & 0.837 ± 0.122          \\
BrainIB ($K$ = 4)       & 0.680 ± 0.037          & 0.674 ± 0.020          & 0.896 ± 0.054          \\ \bottomrule
\end{tabular}
\label{tab:edges}
\end{table}

\section{Conclusion}

In this paper, we develop BrainIB, a novel GNN framework based on information bottleneck (IB) for interpretable psychiatric diagnosis. To the best of our knowledge, this is the first work that uses IB principle for brain network analysis. BrainIB is able to effectively recognize disease-specific prominent brain network connections and demonstrates superior out-of-distribution generalization performance when compared against baselines and state-of-the-art (SOTA) methods on  three challenging psychiatric prediction datasets. We also validate the rationality of our discovered biomarkers with clinical and neuroimaging finds. Our improvements on generalization and interpretability make BrainIB  moving one step ahead towards practical clinical applications. In the future, we will consider combining BrainIB with multimodal data to further improve the accuracy of psychiatric diagnosis and we collect additional datasets to serve as validation sets ($n >1000$). 



\bibliographystyle{IEEEtran}
\bibliography{TNNLS.bib}

\end{document}